\documentclass[fleqn,usenatbib]{mnras}
\usepackage[T1]{fontenc}
\usepackage{ae,aecompl}
\usepackage{graphicx}
\usepackage{amsmath}
\usepackage{amssymb}
\usepackage{url}
\graphicspath{{figure/}}

\newcommand{\kms}{\,km\,s$^{-1}$}
\newcommand{\molh}{$\text{H}_2$}
\newcommand{\Msun}{M$_\odot$}
\newcommand{\pc}[2]{pc$^{-2}$}
\newcommand{\angstrom}{\mbox{\normalfont\AA}}

\defcitealias{Somerville2008}{S08} 
\defcitealias{Somerville2012}{S12}
\defcitealias{Popping2014}{PST14}
\defcitealias{Somerville2015}{SPT15}
\defcitealias{Somerville2015a}{SD15}
\defcitealias{Naab2017}{NO17}

\defcitealias{Kennicutt1998}{KS}
\defcitealias{Gnedin2011}{GK}
\defcitealias{Bigiel2008}{Big}

\title[SAM forecasts - I. UV luminosity functions]{Semi-analytic forecasts for \textit{JWST} - I. UV luminosity functions at $\boldsymbol{z}$ = 4 - 10}

\author[L. Y. A. Yung et al.]{L. Y. Aaron Yung$^{1}$\thanks{E-mail: yung@physics.rutgers.edu},
Rachel S. Somerville$^{1,2}$, Steven L. Finkelstein$^{3}$, 
\newauthor Gerg\"{o} Popping$^{4}$, Romeel Dav\'{e}$^{5,6,7}$
\\
$^{1}$ Department of Physics and Astronomy, Rutgers University, 136 Frelinghuysen Road, Piscataway, NJ 08854, USA\\
$^{2}$ Center for Computational Astrophysics, Flatiron Institute, 162 5th Ave, New York, NY 10010, USA\\
$^{3}$ Department of Astronomy, The University of Texas at Austin, Austin,
TX 78712, USA\\
$^{4}$ Max-Planck-Institut f\"{u}r
Astronomie, K\"{o}nigstuhl 17, 69117 Heidelberg, Germany\\
$^{5}$ Institute for Astronomy, University of Edinburgh, EH9 3HJ, UK\\
$^{6}$ University of the Western Cape, Cape Town 7535, South Africa\\
$^{7}$ South African Astronomical Observatory, Cape Town 7925, South Africa
}

\date{Accepted 2018 November 19. Received 2018 October 24; in original form 2018 March 26}

\pubyear{2018}

\begin{document}
\label{firstpage}
\pagerange{\pageref{firstpage}--\pageref{lastpage}}
\maketitle

\begin{abstract}
In anticipation of the upcoming deployment of the \textit{James Webb Space Telescope} (\textit{JWST}), we present high-redshift predictions by the well-established Santa Cruz semi-analytic model. We update the models by re-calibrating them after adopting cosmological parameters consistent with recent constraints from Planck. We provide predictions for rest-frame UV luminosity functions for galaxy populations over a wide range of $M_\text{UV}$ from $\sim-6$ to $\sim-24$ between $z = 4$ -- 10. In addition, we present the corresponding predictions for observed-frame galaxy number counts in different redshift bins in the full set of NIRCam filters. We provide predictions of the quantitative effect on these observables of varying the physical recipes implemented in the models, such as the molecular gas depletion time (star formation efficiency) scalings or the scalings of outflow rates driven by stars and supernovae with galaxy circular velocity. Based on these results, we discuss what may be learned about the physical processes that shape galaxy formation from \textit{JWST} observations of galaxy number densities at different intrinsic luminosities. All data tables for the results presented in this work are available at \url{https://www.simonsfoundation.org/semi-analytic-forecasts-for-jwst/}.
\end{abstract}

\begin{keywords}
galaxies: evolution -- galaxies: formation -- galaxies: high-redshifts -- galaxies: star formation -- galaxies: statistics
\end{keywords}

\section{Introduction}

The soon-to-be-launched \textit{James Webb Space Telescope} (\textit{JWST}) will possess the unprecedented infrared sensitivity and spatial resolution required for detecting faint, distant galaxies that are extremely difficult or impossible to detect with any current facilities. These observations will provide significant insights into the statistical properties of the galaxy population near cosmic dawn. This is of great interest for determining whether the physical processes that shape galaxy formation are very different in the early universe from locally, as well as for constraining which objects are responsible for reionizing the Universe. 

\textit{Hubble Space Telescope} (\textit{HST}) has detected nearly 2000 galaxy candidates at high redshifts ($z \sim 4$ -- 10) from both blank and gravitationally lensed fields \citep{Koekemoer2013,Lotz2017}. Additionally, brighter objects have also been discovered with ground-based facilities, such as the United Kingdom Infra-Red Telescope (UKIRT) and the Visible and Infrared Survey Telescope for Astronomy (VISTA) \citep{McLure2009,Bowler2015}. These observations have provided constraints on the space density of relatively bright galaxy populations up to $z \sim 10$ \citep[e.g.][]{McLure2009, McLure2013, Castellano2010, VanderBurg2010, Oesch2013, Oesch2014, Oesch2018, Schenker2013, Tilvi2013, Bowler2014, Bowler2015, Bouwens2014, Bouwens2015, Bouwens2016, Bouwens2017, Schmidt2014, Atek2015,  McLeod2015, McLeod2016, Finkelstein2015, Livermore2017, Ishigaki2018}. However, the constraints on the faintest populations currently rely solely on fields that are lensed by massive foreground galaxy clusters, and the corrections for magnification are quite uncertain \citep[e.g.][]{Kawamata2016,Bouwens2017,Priewe2017}.
\textit{JWST} will probe much deeper down the luminosity function in unlensed fields, as well as providing more secure redshift measurements for high-$z$ candidates that are currently selected by the Lyman-break technique. Many members of the community are currently engaged in planning the optimal strategies to take advantage of the limited lifetime of \textit{JWST} to achieve these breakthroughs in our understanding of the very early Universe. Theoretical predictions may be able to aid in designing the observing strategy and trade-offs in imaging area, depth, and wavelength coverage. Perhaps more importantly, it is critical to assess the \emph{uncertainties} on current theoretical predictions and the plausible range in galaxy properties at these extreme redshifts.

With the cosmological parameters that govern early structure formation fairly well constrained by experiments such as the Wilkinson Microwave Anisotropy Probe \citep[WMAP;][]{Spergel2003,Spergel2007,Komatsu2009,Komatsu2011,Hinshaw2013} and Planck satellite \citep{Planck2014,Planck2016}, combined with other constraints from Baryon Acoustic Oscillations, Supernovae, and weak lensing \citep{March2011,Aubourg2015,Hildebrandt2017}, the major uncertainties in forecasting the abundance of galaxies in the early Universe arise from our lack of a rigorous theory for how dense, cold molecular clouds form, how stars form within these clouds, and how thermal energy, momentum, and radiation from stars and supernovae affect the subsequent efficiency of star formation (``stellar feedback''; see \citet{McKee2007}, \citet[hereafter \citetalias{Somerville2015a}]{Somerville2015a}, and \citet[hereafter \citetalias{Naab2017}]{Naab2017}). Although feedback from radiation and jets produced by accretion onto supermassive black holes is likely to be critical for regulating galaxy formation in the lower redshift Universe (\citetalias{Somerville2015a} and references therein), at very early times $z \gtrsim 6$ most galaxies probably have not had time to form massive black holes in their nuclei, so it is likely (although not certain), that this form of feedback is sub-dominant at these epochs. 

Some theorists have suggested that star formation remains very efficient out to very high redshifts \citep[e.g.][]{Behroozi2015}, while others have suggested that the low metallicity environments will make it difficult to form molecular hydrogen and lead to inefficient star formation \citep[e.g.][]{Krumholz2012}. The efficiency of star formation in low-mass halos at early times is also critical to determining which objects reionized the Universe, and how reionization proceeded in space and time. Numerous studies have shown that the already detected galaxy population is likely insufficient to reionize the Universe by $z\sim 6$--10 as required by observations \citep{Finkelstein2015, Robertson2015}, and it is generally assumed that the shortfall in the ionizing photon budget is made up by faint galaxies that form in low-mass halos \citep{Bouwens2012, Kuhlen2012, Atek2015, Gnedin2016, Anderson2017}. Many published theoretical predictions for reionization assume a fixed efficiency for converting baryons into stars in halos \citep{Vogelsberger2013,Hopkins2014,Mutch2016}. However, it is known that in the very nearby Universe, star formation is extremely inefficient in low-mass halos, leading to questions about whether the local and high redshift observations are in tension. 

Computational methods have long been used in modeling the formation and evolution of galaxies. However, given that these processes operate over a vast range of scales, both spatially and temporally, modeling galaxy formation in a cosmological context remains one of the greatest challenges in astrophysics today \citepalias{Somerville2015a}. One possible approach to predicting galaxy properties at high redshift is to constrain the relationship between galaxy and dark matter halo properties at lower redshift, and then assume that these relationships hold out to higher redshift \citep[e.g.][]{Behroozi2015,Mason2015,Furlanetto2017}. Another is to implement physical recipes for the key processes that are thought to shape galaxy formation (e.g. gas accretion and cooling, star formation and stellar feedback, chemical enrichment, black hole formation and feedback) within a cosmological framework (see \citetalias{Somerville2015a}, \citetalias{Naab2017} for a recent review). These physical recipes may be implemented within numerical simulations, which attempt to explicitly follow the equations of gravity, hydrodynamics, and thermodynamics for particles or grid cells \citep{Springel2001, Bryan1997a, Bryan1999}. However, current cosmological simulations are unable to directly resolve the multiphase structure of the interstellar medium, the formation and evolution of individual stars and supermassive black hole (SMBH), or their interaction with their surroundings. Thus, ``sub-grid'' recipes must be used to represent these processes on scales below those that can be explicitly resolved in the simulations. In most current numerical simulations, these sub-grid recipes are generally phenomenological and contain free parameters that must be calibrated to match global galaxy observations (see discussion in \citetalias{Somerville2015a}). Due to the high computational expense, rather limited dynamic range can be achieved with these techniques, and the ability to explore different sub-grid recipes and parameter values is also limited.

An alternative is semi-analytic simulations, which apply simplified recipes for these same physical processes within dark-matter halo ``merger trees'' either extracted from dissipationless N-body simulations or created using semi-analytic Monte Carlo techniques. Despite the absence of a spatially defined ``grid'' in semi-analytic models (SAMs), the essence of using phenomenological or physically driven recipes to model galaxies is very similar to the motivation of sub-grid recipes in conventional numerical simulations. One way to view SAMs is as, essentially, a book-keeping scheme that tracks the movement of mass between different reservoirs, as gas accretes from the diffuse intergalactic medium (IGM) into galactic halos, flows from halos into the interstellar medium (ISM) of galaxies, is converted from ISM into stars, and potentially is ejected from the ISM back into the hot diffuse galactic halo or back out into the IGM. In addition to gas and stars, heavy elements can also be tracked as they are produced by stars and circulate through the cosmic baryon cycle. Like the numerical simulations, the physical recipes in SAMs are phenomenological, and contain free parameters that must be calibrated to global galaxy observations. 

This method has been widely used to explore a very broad range of galaxy properties \citep{White1991, Kauffmann1993a, Cole1994, Somerville1999, Croton2006, Mutch2016} and it has shown that the predictions for the statistical properties of galaxies obtained using these methods are in very good agreement with those from numerical simulations (\citetalias{Somerville2015a}). Due to the much greater computational efficiency and flexibility of these methods, however, they provide several advantages over the numerical approach, namely by allowing researchers to explore a broader range of parameter space and different physical recipes. Moreover, SAMs are also able to span a larger dynamical range in dark matter halo mass, from the smallest halos that are believed to be able to form stars to the most massive and rare objects. 

Recently, many other groups have made theoretical predictions for high-$z$ \textit{JWST} observations using numerical (e.g. \citealt{Barrow2017}), phenomenological(e.g. \citealt{Williams2018}), and semi-analytic (e.g. \citealt{Cowley2018}) methods. These works collectively probe a broad range of aspects, including the physical and photometric properties of high-$z$ galaxies \citep{Cowley2018}, population synthesis and synthetic spectra \citep{Barrow2017, Volonteri2017}, and the impact of broadband filter choice on estimating photometric redshifts and on recovering their physical properties from observations \citep{Bisigello2016, Bisigello2017}. Despite the great interest in using \textit{JWST} to uncover the physical nature and the assembly histories of high-$z$ galaxies, however, only a handful of studies have systematically varied the underlying physics or model parameters in a controlled way within the same modeling framework. We find that these kinds of predictions will be quite informative for the interpretation of \textit{JWST} observations.

In this work, we use the well-established Santa Cruz SAM, which has been shown to successfully reproduce key galaxy observations at lower redshifts ($z\lesssim6$), to make predictions for galaxy populations at intermediate to high redshifts ($z = 4$--10). Importantly, we make use of the Santa Cruz SAM version that tracks multiple phases of gas in the ISM (atomic, molecular, and ionized) using a suite of recipes based on empirical considerations or predictions from detailed numerical simulations containing treatments of molecular chemistry and radiative transfer (\citealt{Popping2014}, hereafter \citetalias{Popping2014}; \citealt{Somerville2015}, hereafter \citetalias{Somerville2015}). Our fiducial model is based on the empirically grounded assumptions that stars form in environments that are dominated by molecular gas, and that the formation of molecular gas depends on the gas surface density, metallicity, and the local UV radiation field. However, we also explore how sensitive our predictions at ultra-high redshift are to our recipes for star formation and stellar feedback. In this paper we focus on first-order, directly observable quantities, namely one-point distribution functions for rest-frame UV luminosity and observed-frame magnitudes in \textit{JWST} NIRCam filters. In a series of planned papers, we will explore physical properties of galaxies such as stellar masses, molecular gas content, metallicities, and dark matter halo masses, and will also make direct predictions for the implications for reionization. 

The key components of this work are summarized as follows: the basic elements of the Santa Cruz SAM used in this work are summarized in \S\ref{section:sam}. We then present the rest-frame UV LFs in \S\ref{section:uvlf} and counts in observed \textit{JWST} filters in \S\ref{section:nircam}. We discuss our results in \S\ref{section:discussion}, and summarize and conclude in \S\ref{section:summary}.

\section{The Semi-Analytic Framework}
\label{section:sam}

The SAM used in this study is the same one outlined in \citet[hereafter \citetalias{Somerville2015}]{Somerville2015}. Hence, we will not describe the model in full here and we refer the reader to the following works for full detail for the  modeling framework developed by the Santa Cruz group: \citet{Somerville1999}; \citet*{Somerville2001}; \citet{Somerville2008, Somerville2012}; \citetalias{Popping2014} and \citetalias{Somerville2015}. Throughout this work, we adopt cosmological parameters that are consistent with the ones reported by \citeauthor{Planck2016} in 2015: $\Omega_\text{m} = 0.308$, $\Omega_\Lambda = 0.692$, $H_0 = 67.8$\kms Mpc$^{-1}$, $\sigma_8 = 0.831$, and $n_s = 0.9665$.

Dark matter halo merger histories, more commonly known as `merger trees', are the backbone of the semi-analytic modeling framework. These merger trees can either be extracted from dissipationless $N$-body simulations or constructed using semi-analytic methods based on the Extended Press-Schechter (EPS) formalism \citep{Press1974, Lacey1993}. In this work, in order to maximize the dynamic range and computational efficiency, we adopt the EPS-based method of \citet{Somerville1999b} with updates as described in \citet[hereafter \citetalias{Somerville2008}]{Somerville2008}. We have also run our models using halo merger trees from the Bolshoi Planck $N$-body simulations, and find  very similar results over the dynamical range spanned by that simulation. However, in addition to having inadequate mass resolution to resolve the low-mass halos that host the faint galaxy population that will be detectable by JWST, Bolshoi Planck has only 29 snapshots stored above $z=6$, which is inadequate to construct accurate merger histories for $z>6$ galaxies. In addition, robust identification of halos at high redshift becomes tricky. Friends-of-friends methods, such as used in e.g. the \textsc{BlueTides} simulation \citep{Feng2015} and the DRAGONS simulation suite \citep{Poole2016}, have been shown to artificially link together distinct halos at high redshift \citep{Klypin2011}. Although we are in the process of generating a new suite of simulations, halo catalogs, and merger trees that are carefully designed to model faint galaxies in the high redshift Universe (Yung et al. in prep), we are confident that the EPS-based approach has sufficient accuracy for the somewhat qualitative exploration that we present here.

At each output redshift, we set up a grid of one hundred root halos with masses spanning the range in virial velocity $V_\text{vir} \approx 20 - 500$\kms, which covers halos ranging from close to the atomic cooling limit to the rarest objects expected to be detected in high-redshift surveys. We then weight each of these root halos by the expected abundance of dark matter halos of the given mass at the respective redshift, using the fitting functions  provided in \citet{Rodriguez-Puebla2016} based on results from the MultiDark suite \citep{Klypin2016} of $N$-body simulations. Although this involves extrapolating the fitting functions to lower halo masses and higher redshifts than those that are directly probed by the MultiDark Suite, we have have validated these results using an unpublished suite of very high resolution, small box simulations kindly made available to us by Eli Visbal \citep{Visbal2018}. The assembly history is then traced down to a minimum progenitor mass of $M_\text{res}$, which we refer to as the mass resolution of our simulations; here we set $M_\text{res} = 10^{10}$\Msun\ or 1/100th of the root halo mass, whichever is smaller. For each root halo in the grid, one hundred Monte Carlo realizations of the merger histories are generated.

After constructing these semi-analytic merger trees, our SAM implements a suite of fairly standard recipes such as cosmological accretion and cooling, star formation and stellar-driven winds, chemical evolution, black hole feedback, and  mergers. We again refer readers to \citetalias{Somerville2008} and \citetalias{Somerville2015} for full details.

\subsection{Gas Partitioning and Star Formation}

In the most recent iteration of the Santa Cruz SAM (\citetalias{Popping2014}, \citetalias{Somerville2015}), disks are subdivided into annuli and the cold gas in each annulus is partitioned into an atomic (\ion{H}{I}), ionized (\ion{H}{II}), and molecular (\molh) component. In \citetalias{Popping2014} and \citetalias{Somerville2015}, several different recipes for gas partitioning are investigated, including an empirical recipe in which the molecular fraction is determined by the disk mid-plane pressure, and several variants of recipes in which the molecular fraction is determined by the gas surface density, metallicity, and the intensity of the local UV radiation field. Somewhat surprisingly, these studies found that most results were not very sensitive to which gas partitioning recipe was used. However, they did find that the recipe based on the prescription of \citet{Krumholz2009}, in which the dependence on the UV background was not taken into account, failed to reproduce sufficient numbers of low-mass galaxies and was disfavored. Overall, the metallicity and UV-background dependent recipe based on simulations by \citet[hereafter \citetalias{Gnedin2011}]{Gnedin2011} was found to perform the best, and was adopted as the ``fiducial'' model. As we are concerned that some assumptions contained in the empirical pressure-based model may no longer hold at high redshift, we do not explore it here.

The usual picture is that the first stars form out of ``primordial'' molecular hydrogen, form Pop III stars, and pollute early halos with metals \citep{Frebel2009, Wise2012}. In the current work we do not model the formation of metal-free Population III  stars explicitly. Instead (as in previous works), we set a metallicity floor of  $Z_\text{pre-enrich}$ to represent pre-enrichment from Pop III stars. We adopt $Z_\text{pre-enrich} = 10^{-3}Z_\odot$ \citep{Bromm1999}. In SPT15, we show that our results do not depend sensitively on the choice of this value.

Many earlier generations of SAMs adopt what we refer to as the ``classic'' Kennicutt-Schmidt (\citetalias{Kennicutt1998}) relation to model the rate at which cold gas is converted into stars. In this approach, the star formation rate (SFR) is assumed to scale as a power of the total (cold; \ion{H}{I}+\molh) gas density \citep{Schmidt1959, Schmidt1963, Kennicutt1989, Kennicutt1998}, where in some implementations (as in \citetalias{Somerville2008}), only gas above a fixed surface density is assumed to participate in star formation. This may approximate to first order the transition from predominantly molecular to predominantly atomic gas that the above multiphase modeling attempts to capture, but there is observational evidence that this critical surface density depends on gas metallicity \citep{Wilson1995, Arimoto1996, Bolatto2008, Bolatto2011, Genzel2010, Leroy2011}.

A more recent generation of SAMs adopts a \molh-based star formation recipe \citep{Lagos2011, Fu2012, Fu2013, Somerville2015, Xie2017}. Observations of nearby spirals have shown that the SFR surface density is nearly linearly proportional to the molecular hydrogen surface density \citep{Wong2002, Bigiel2008, Bigiel2011, Leroy2011}. However, these observations only probe \molh\ surface densities up to about 50-80 \Msun\pc2\ . Mounting evidence from both observation and theory suggests that the slope of the SF relation may steepen to $\sim 2$ at higher surface densities \citep{Sharon2013, Rawle2014, Hodge2015, Tacconi2018}. We have explored both a ``single slope'' SF relation (which we refer to as Big1), in which the molecular gas depletion time is effectively invariant with both galaxy properties and redshift, and a ``two slope'' relation (referred to as Big2), in which the molecular gas depletion time decreases (and star formation efficiency increases) with increasing \molh\ surface density. The surface density of SFR is given by the expression:
\begin{equation}
\Sigma_\text{SFR} = \frac{A_\text{SF}}{\tau_{*,0}}\left(\frac{\Sigma_\text{\molh}}{10\text{M}_\odot\text{pc}^{-2}}\right) \left(1+\frac{{\Sigma_\text{\molh}}}{{\Sigma_\text{\molh,crit}}}\right)^{N_\text{SF}} \text{,}
\end{equation}
where the critical \molh\ surface density $\Sigma_\text{\molh,crit} = 70$ \Msun\pc2\,, $A_\text{SF}$ is the SF relation normalization, and $\tau_{*,0}$ is a tunable normalization parameter. We adopt a value of $A_\text{SF}$ from observational determinations of the relevant Kennicutt-Schmidt relation (as given in Table \ref{table:models}), and allow $\tau_{*,0}$ to vary by about a factor of 50\% up or down, reflecting the observational uncertainty in the true normalization of the Kennicutt-Schmidt relation (although here we find that $\tau_{*,0}=1$ produces good results).  In \citetalias{Somerville2015}, we found that although at $z\sim 0$, the predictions for galaxy properties were not very sensitive to which SF relation was adopted, an increasingly large discrepancy was seen at higher redshifts, up to the highest redshifts $z\sim 6$ explored in that paper. Accordingly, in this work we explore the implications of adopting these different SF recipes at even higher redshifts. The model variants are summarized in Table \ref{table:models}. 

\begin{table}
	\centering
	\caption{Summary of the gas partitioning (GP) and star formation (SF) model variants explored in this work, where $N_\text{SF}$ is the SF relation slope. GKBig2 is our fiducial model.}
	\label{table:models}
	\begin{tabular}{ l  l  l  c  l } 
		\hline
		Model & GP recipe & SF law & $N_\text{SF}$ & $A_\text{SF}$\\
		\hline
		KS & None & \citetalias{Kennicutt1998}  & 1.4 & $1.1\times 10^{-4}$ \\ 
		GKBig1 & \citetalias{Gnedin2011} & \citetalias{Bigiel2008}1  & 1.0 & $4.0\times 10^{-3}$ \\
		GKBig2 (Fid.) & \citetalias{Gnedin2011} & \citetalias{Bigiel2008}2 & $2.0$ & $4.0\times 10^{-3}$ \\
		\hline
	\end{tabular}
\end{table}

\subsection{Photoionization Feedback}
Gas accretion and, ultimately, star formation activity in a galaxy can be reduced in the presence of a strong photoionizing background \citep{Efstathiou1992, Thoul1996, Quinn1996}; this effect is sometimes known as photoionization `squelching' \citep{Somerville2002}. Studies have shown that this effect is especially effective in low-mass halos, and it is thought to be significant in suppressing the collapse of gas into small mass halos and preventing the overproduction of dwarf galaxies in the local universe \citep{Bullock2000, Benson2002, Benson2002a, Somerville2002}. 

As in our previously published models, we adopt the approach proposed by \citet{Gnedin2000} to model photoionization squelching, in which the fraction of baryons that can collapse into halos of a given mass $M_\text{halo}$ at redshift $z$ in the presence of a photoionizing background is computed using the following function:
\begin{equation}
f_\text{b}(M_\text{halo}, z) = \langle f_\text{b}\rangle\left\{1+(2^{\alpha/3} - 1) \left[\frac{M_\text{halo}}{M_\text{char}(z)}\right]^{-\alpha}\right\}^{-3/\alpha} \text{,}
\end{equation}
where $\langle f_\text{b}\rangle$ is the cosmic mean baryon fraction, and $M_\text{char}(z)$ is the mass at which halos retain half of the universal baryon fraction. In the present work, we adopt the redshift dependent characteristic mass obtained by \citet{Okamoto2008} from hydrodynamic simulations including a uniform metagalactic background (based on tabulated results kindly provided to us by T. Okamoto), and adopt $\alpha = 2$ as favored by those authors. We switch squelching on at a fixed redshift  $z_\text{squelch} = 8$, which is consistent with the redshift at which instantaneous reionization occurred as reported by the \citet{Planck2016a}.

The characteristic mass $M_\text{char}$ that we adopt ranges from a value of $1.4 \times 10^7$ \Msun\ just after reionization to a value of $\sim 9.3 \times 10^{9}$ \Msun\ (corresponding to a virial velocity of $\sim 25$ km/s) at $z=0$. Note that these currently favored values of the characteristic mass \citep[][]{Hoeft2006,Okamoto2008} are significantly smaller than the values obtained by \citet{Gnedin2000}, and used in previous versions of the Santa Cruz SAM.

Our treatment of squelching is rather crude and of course is not self-consistent. Some previous semi-analytic studies have attempted to model photoionization squelching in a more self-consistent manner, based on the predicted UV emissivity of the galaxy population that emerges in the SAM \citep{Benson2002, Benson2002a}, and some numerical simulations have also included photoionization squelching self-consistently \citep{Finlator2012, Gnedin2014a}. However, these studies have found that the self-consistent modeling of squelching compared with the simpler approach adopted here does not have a significant effect on predictions for observable galaxy properties or reionization \citep{Mutch2016}. Indeed, as we show in fig. \ref{fig:UVLF}, the \citeauthor{Okamoto2008} filtering mass is so low that squelching has an almost undetectable effect on any observable galaxy properties at the redshifts that we investigate here. This is in agreement with the result found by \citet{Kim2013}.

\subsection{Black Hole Growth and Feedback}
Top-level halos are seeded with black holes with an initial mass of $10^4$\Msun. Black hole accretion and growth may occur in two modes. Mergers and disk instabilities trigger relatively rapid, radiatively efficient accretion (see S08 and \cite{Hirschmann2012a} for details). In addition, a less rapid, radiatively inefficient mode of fueling is associated with Bondi accretion from the hot halo (again, see S08 for details). Two modes of AGN feedback are also implemented, associated with each accretion mode. The radiatively efficient mode drives winds that eject material from the cold gas reservoir and drive it out of the halo. The radiatively inefficient mode is associated with relativistic jets that are assumed to heat the hot halo gas, reducing or even quenching cooling. The two models of AGN feedback have associated parameters, $\epsilon_{\rm wind}$ (see Eqn. 17 of \citetalias{Somerville2008}) and $\kappa_\text{AGN}$ (equivalent to $\kappa_\text{radio}$ in Eqn. 20 of \citetalias{Somerville2008}). The parameter $\epsilon_{\rm wind}=0.5$ is set based on results from hydrodynamic simulations of binary galaxy mergers (see \citetalias{Somerville2008}). The value of the parameter controlling the ``jet mode'' feedback ($\kappa_\text{AGN}$) is adjusted to fit constraints on the relationship between stellar mass and halo mass from abundance matching, and the stellar mass function derived from observations (see Appendix \ref{section:recalibration}).

In practice, we find that the implementation of AGN feedback currently included in our model has no noticeable impact on the predictions presented in this paper -- our results remain unchanged when we switch off both modes of AGN feedback. However, AGN feedback does impact the calibration of the model shown in the Appendix, and so we describe it here for completeness.

\subsection{Stellar Feedback}
Star formation and the baryon fraction in galaxies is thought to be regulated by large-scale outflows driven by massive stars and supernova explosions. The details of how these winds are driven and what determines their efficiency are poorly understood, and cosmological simulations are in general unable to drive these outflows directly without adopting various ``tricks'' (see \citetalias{Somerville2015a} for a discussion). A frequently adopted assumption, loosely motivated by the expectations of momentum-conserving or energy-conserving winds (see \citetalias{Somerville2015a}) is that the mass outflow rate scales with a power of the galaxy potential well depth, represented by the internal velocity dispersion or rotation velocity, times the star formation rate:

\begin{equation}
\dot{m}_\text{out} = \epsilon_\text{SN} \left( \frac{V_0}{V_\text{c}}\right)^{\alpha_\text{rh}} \dot{m}_*\text{,}
\end{equation}
where $\dot{m}_*$ is the star formation rate, $V_\text{c}$ is the circular velocity of the galaxy, normalized by the arbitrary constant $V_0 = 200$\kms, and  $\epsilon_\text{SN}$ and $\alpha_\text{rh}$ are tunable free parameters.

Variants of this recipe are almost universal in semi-analytic models, and several numerical cosmological hydrodynamic simulations effectively adopt it by administering ``kicks'' to particles according to these or similar scaling relations \citep{Oppenheimer2006, Vogelsberger2014, Dave2016}. Intriguingly, several simulations in which winds are driven by attempting to implement stellar feedback ``directly'' also find that the wind mass loading factors $\eta \equiv  \dot{m}_\text{out}/\dot{m}_*$ follow this type of scaling. \citet{Christensen2016} find that their zoom-in simulations with the GASOLINE code and a ``blastwave'' approach to driving winds yield $\alpha_{\rm rh} \sim 2$, and \citet{Muratov2015} find that at high redshift $\eta$ depends strongly on $V_c$ in the FIRE simulations, where $\alpha_{\rm rh} \gtrsim 2$ for low-$V_{\rm c}$ halos and $\alpha_{\rm rh} \sim 1$ at higher $V_{\rm c}$.

\subsection{Stellar Populations and Dust Attenuation}

For each galaxy, we store a two-dimensional grid recording the mass of stars formed with a given age and metallicity. Unattenuated synthetic spectral energy distributions (SED) are created for each galaxy by convolving this grid with simple stellar population models (SSP). Here we use the SSP models of \citet{Bruzual2003} with the Padova1994 \citep{Bertelli1994} isochrones and a \citet{Chabrier2003} initial mass function (IMF). 

We include dust attenuation using a simple approach similar to the one presented in \citet[hereafter \citetalias{Somerville2012}]{Somerville2012}. We assume that the face-on extinction optical depth in the $V$-band is given by
\begin{equation}
\tau_{V,0} = \tau_{\text{dust,}0}\,Z_\text{cold}\,m_\text{cold}/(r_\text{gas})^2 \text{,}
\label{eqn:opticaldepth}
\end{equation}
where the optical depth normalization of dust, $\tau_{\text{dust},0}$, is a free parameter, $Z_\text{cold}$ is the metallicity of the cold gas, $m_\text{cold}$ is the mass of the cold gas in the disc, and $r_\text{gas}$ is the radius of the cold gas disc. We then calculate  the attenuation in the $V$-band for a given inclination using a ``slab'' model, which assumes that the radiation sources (stars) are embedded in a slab of dust:
\begin{equation}
A_V = -2.5\log_{10}\left[\frac{1-\exp[-\tau_{V,0}/\cos(i)]}{\tau_{V,0}/\cos(i)}\right] \text{.}
\label{eqn:dust}
\end{equation}
Here, $i$ is the inclination, which is chosen randomly for each galaxy. We then translated this to the attenuation in the UV-band (1600\angstrom), $A_\text{UV}$, assuming a fixed attenuation curve \citep{Calzetti2000}. We experimented with using the two-component \citet{Charlot2000}-type model (as implemented in \citetalias{Somerville2012}), in which an age-dependent attenuation curve is effectively adopted by assuming that young stars are enshrouded in higher optical-depth ``birth clouds''. However, we found that this type of model predicted very large amounts of extinction in high redshift galaxies, which was incompatible with observations. 

\begin{figure}
	\includegraphics[width=\columnwidth]{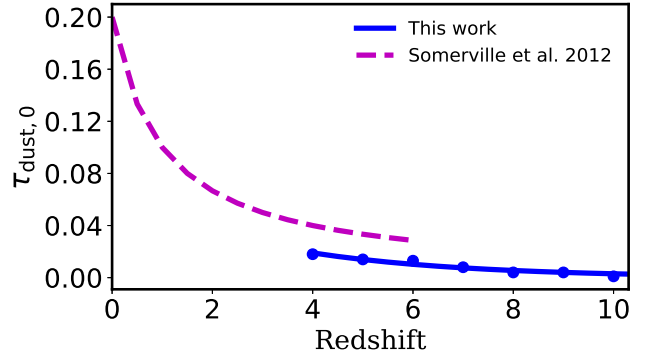}
	\caption{Comparison of the $\tau_\text{dust,0}$ values adopted in this work and in \citetalias{Somerville2012}. The blue markers show the values tuned by hand to match observations, and the solid line is the fitting function adopted in our model (see text).}
	\label{fig:tau_dust_0} 
\end{figure}

Several previous works have found that simple models for dust attenuation, in which the dust extinction optical depth scales with gas metallicity and column density with a fixed normalization, overpredict the amount of attenuation in high redshift galaxies \citep[\citetalias{Somerville2012};][]{Wilkins2013, Reddy2015, Reddy2018, Whitaker2017a}. One possible interpretation is that the dust-to-metal ratio is not constant, and depends on galaxy properties that effectively cause it to change with cosmic time. Indeed, recent observations have shown that the dust-to-metal ratio differs significantly across different galaxy mass scales \citep{Remy-Ruyer2014}, and models that attempt to self-consistently track the production and destruction of dust by various channels find that the dust-to-metal ratio may change over cosmic time \citep{Popping2017}. We plan to develop more detailed and physical models of dust attenuation in high redshift galaxies in future work (Popping et al. in prep), but for the present work we adopt a very simple and ad hoc approach in which we allow the dust normalization parameter $\tau_\text{dust,0}$ to be a function of redshift \citep[\citetalias{Somerville2012}, see also:][]{Inoue2003,Guo2009,LoFaro2009,Santini2014,Wiseman2017}. We tune $\tau_\text{dust,0}$ by hand and find the values summarized in Table \ref{table:tau_dust_0} provide a good match to available observational constraints on the bright-end of the UV LFs at $z = 4$ -- 10. This is then implemented in our model with a functional form $\tau_\text{dust,0} = \exp(az+b)$ with parameter values $a = -0.31$ and $b = -2.726$. A comparison of the values adopted in this work and in \citetalias{Somerville2012} can be found in fig. \ref{fig:tau_dust_0}. Note that the cosmological parameters and SAM ingredients and parameters adopted in \citetalias{Somerville2012} are a little bit different than the ones adopted in this work, which is why the values do not match up perfectly in the redshift range of overlap.

\begin{table}
	\centering
	\caption{Values for $\tau_\text{dust,0}$ tuned by hand to match observations and fitted with the exponential function provided in the text.}
	\label{table:tau_dust_0}
	\begin{tabular}{ @{\hspace{4em}}c @{\hspace{2.5em}}c@{\hspace{1.5em}} c@{\hspace{4em}} } 
		\hline
        & \multicolumn{2}{c@{\hspace{4em}}}{$\tau_{\text{dust},0}\;(10^{-3})$} \\ [0.5ex] 
		$z$ &  Tuned  &  Fitted \\ [0.5ex] 
		\hline
		4 & 18.00 & 18.99 \\ 
		5 & 14.00 & 13.93 \\
		6 & 13.00 & 10.22 \\ 
		7 & 8.00  & 7.50 \\
		8 & 4.00  & 5.50 \\
        9 & 4.00  & 4.04 \\
        10 & 1.00 & 2.96 \\
		\hline
	\end{tabular}
\end{table}

We obtained the latest version of the published NIRCam filter response functions\footnote{\url{https://jwst-docs.stsci.edu/display/JTI/NIRCam+Filters}} with optical telescope element (OTE). For each galaxy, we calculate its apparent magnitude by convolving the synthesized SED with the NIRCam filters. We also include absorption due to the intervening IGM. The effective optical depth of the IGM along the line-of-sight, to a source at some redshift, at wavelength $\lambda$ is calculated using the expression given in \citet{Madau1996}.

\begin{table*}
	\centering
	\caption{A table for the model parameters that changed after re-calibration for the Planck cosmology. We also show the values used in \citetalias{Somerville2015}, where cosmological parameters from WMAP5 were adopted, to illustrate by how much these parameters have changed. For a complete list of model parameters, see Table 1 in \citetalias{Somerville2015}.}
	\label{table:parameters}
	\begin{tabular}{ l  l  c c } 
		\hline
		Parameter       & Description              & This work  & SPT15 \\ [0.5ex] 
		\hline
		$\epsilon_\text{SN}$ &  SN feedback efficiency  & 1.7 & 1.5 \\ 
		$\alpha_\text{rh}$ & SN feedback slope & 2.8 & 2.2 \\
		$\tau_{*,0}$ &  SF timescale normalization & 1.0 & 1.0 \\ 
		$y$ & Chemical yield (in solar units) & 2.0 & 1.6 \\
		$\kappa_\text{AGN}$ & Radio mode AGN feedback & $3.0\times 10^{-3}$ & $3.8\times 10^{-3}$ \\
        \hline
	\end{tabular}
\end{table*}

\subsection{Chemical Evolution} 
The production of metals is modeled using a simple approach that is commonly adopted in semi-analytic models (see e.g. \citealt{Somerville1999}; \citealt{Cole2000}; \citealt{Lucia2004}). In a given timestep where a parcel of new stars $d m_*$ is created, a mass of metals $dM_{Z} = y\,dm_*$ is also formed, where $y$ is the `effective' chemical yield, or mean mass of metals produced per mass of stars. Here we assume that the chemical yield is constant. In principle, $y$ could be obtained from stellar evolution models, but these model yields are uncertain by a factor of $\sim2$, and the single-element instantaneous recycling approach to chemical evolution that we are using here is somewhat crude, so we instead treat the chemical yield as a free parameter while restricting it to an expected range. Once created, metals are assumed to be mixed instantaneously with the cold gas in the disk. We track the mean metallicity of the cold gas $Z_\text{cold}$, and new star parcels created out of this gas are assumed to have the same stellar metallicity $Z_*$ as the mean metallicity of the cold gas in that timestep. Supernova feedback ejects metals from the disk, along with cold gas. These metals are either mixed with the hot gas in the halo, or ejected from the halo into the ``diffuse'' Intergalactic Medium (IGM), in the same proportion as the reheated cold gas(see \citetalias{Somerville2008}). The ejected metals in the `diffuse gas' reservoir are also reaccreted into the halo in the same manner as the gas.

Throughout this paper, the yield $y$ and all metallicities are given in solar units, which we take to be $Z_\odot$ = 0.02. Although this formally represents the total metallicity, we note that as we track only the enrichment associated with Type II supernovae, our metallicity estimates probably correspond more closely with $\alpha$-type elements. Note that because enriched gas may be ejected from the halo, and primordial gas is constantly being accreted by the halo, this approach is not equivalent to a standard ``closed box'' model of chemical evolution.

\subsection{Calibrating the free parameters}
We calibrate our models to a standard set of $z\sim 0$ observables, and then leave all free parameters (except the dust normalization, as noted above) fixed. Relative to the WMAP5 cosmology used in \citetalias{Somerville2008} and \citetalias{Somerville2015}, the Planck cosmology adopted here results in significantly different predictions for the abundance of dark matter halos as a function of cosmic time (see \citealt{Rodriguez-Puebla2016}), and as a result the free parameters of the SAM need to be re-calibrated. We show the calibration quantities, along with some other diagnostic quantities that are not used in calibration, in Appendix \ref{section:recalibration}. The details of the calibration procedure are also presented in the Appendix. The model parameters used throughout this work are summarized in Table \ref{table:parameters}.

\section{Rest-frame UV Luminosity Functions}
\label{section:uvlf}

Perhaps the most basic statistical characterization of the galaxy population is the one-point distribution function of an observable quantity, such as luminosity (commonly referred to as the ``luminosity function''  (LF)). The change of the LF in different redshift bins constrains the evolution of the galaxy population over cosmic time. In this section, we present the UV LFs predicted by our SAM at $z = 4 - 10$. With the large dynamic range of dark matter halo masses probed by our models, our predictions cover a wide UV luminosity range between $M_\text{UV} \sim -6$ to $-24$. All binned luminosity functions presented in this work are available for download online\footnote{ \url{https://www.simonsfoundation.org/semi-analytic-forecasts-for-jwst/}}. Throughout this work we use a tophat filter of width of 400\AA\ centered at 1600\AA\ to calculate the rest-frame UV luminosity.

\begin{figure}
	\includegraphics[width=\columnwidth]{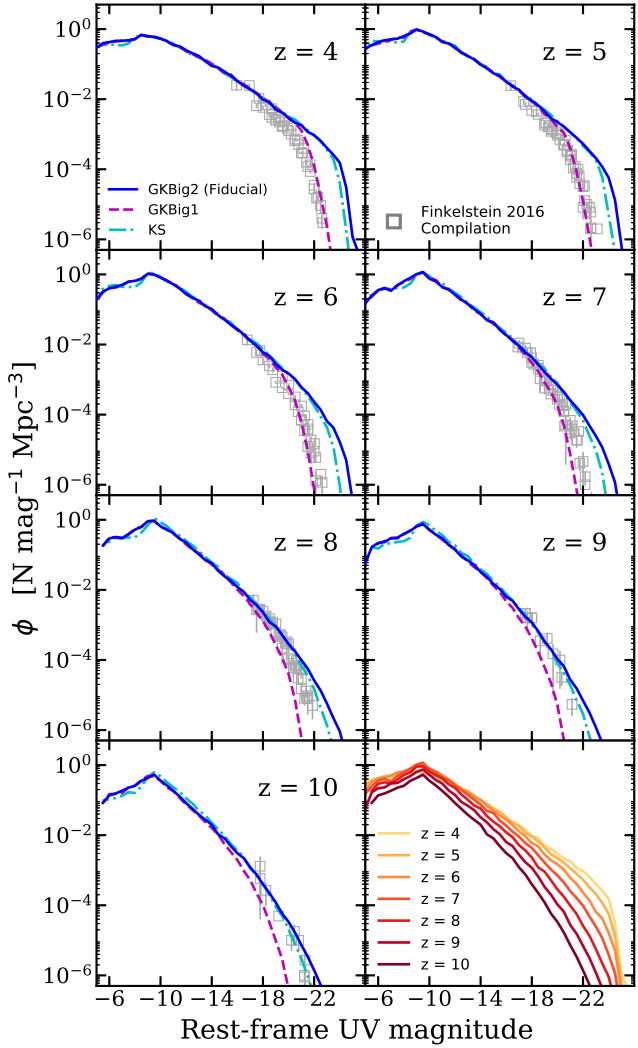}
	\caption{Predicted intrinsic UV LFs (without correction for dust attenuation) and their evolution with redshift. The blue solid line shows the results of the GKBig2 (fiducial) model, the purple dashed line shows the GKBig1 model, and the cyan dot-dashed line shows the KS model. We also include a compilation of observational constraints from \citet[squares]{Finkelstein2016} to guide the eye. The last panel summarizes the evolution predicted by the fiducial model.}
	\label{fig:UVLF_intr} 
\end{figure}

\begin{figure}
	\includegraphics[width=\columnwidth]{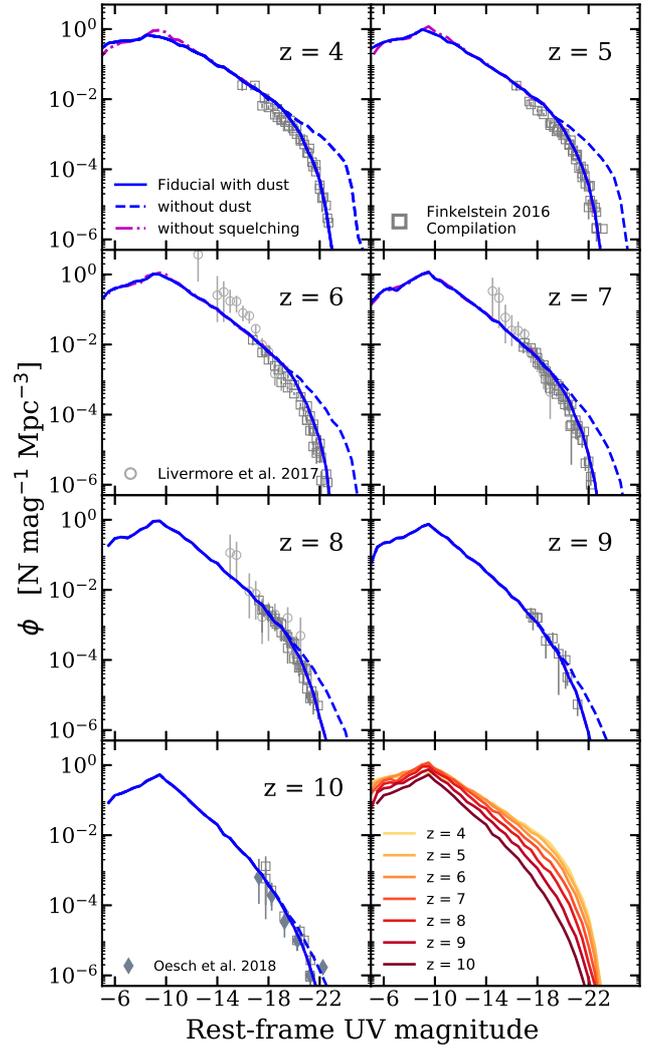}
	\caption{Redshift evolution of the dust-attenuated UV LFs between $z = 4 - 10$ predicted by our fiducial model (blue solid line). The blue dashed line shows the intrinsic UV LFs and the purple dot-dashed line shows the UV LFs  without the effect of photoionization squelching, both from the fiducial model. We also include a compilation of observational constraints from \citet[squares]{Finkelstein2016} (same as fig. \ref{fig:UVLF_intr}) to guide the eye. Addition observational constraints from \citet[circles]{Livermore2017} and \citet[diamonds]{Oesch2018} are shown in $z = 6$, 7, 8, and 10. The last panel summarizes the evolution of the dust-attenuated UV LFs predicted by the fiducial model.}
	\label{fig:UVLF} 
\end{figure}

In fig. \ref{fig:UVLF_intr}, we show the distribution functions for the intrinsic rest-frame UV luminosities, without accounting for the effect of dust attenuation. Our results show that the choice of star formation recipe can significantly alter the number density of bright galaxies. Recall that in the GK-Big1 model, the star formation efficiency (molecular gas depletion timescale) is effectively fixed at a constant value. Keeping in mind that the \molh\ depletion timescale in nearby spirals is 1--2 Gyr \citep{Bigiel2008,Leroy2008}, it is perhaps unsurprising that this becomes the limiting factor in forming stars at times when the age of the Universe is significantly less than this. In the GK-Big2 model, the ``steepening'' of $N_\text{SF} \rightarrow 2$ effectively leads to a higher star formation efficiency and shorter $t_{\rm dep, mol}$ in higher density gas. As galaxies are much more compact and gas rich at high redshift, this effectively leads to higher SF efficiencies at high redshift. The super-linear dependence of the ``classic'' Kennicutt-Schmidt SF relation (\citetalias{Kennicutt1998}) goes in the same direction, but to a lesser extent, as it assumes a slightly shallower slope $N_{\rm SF}=1.5$. Interestingly, as already suggested in \citetalias{Somerville2015}, we find that the formation of molecular gas is not a significant limiting factor for star formation even at these very high redshifts, in contrast to the suggestions of \citet{Krumholz2012}. On the other hand, the faint end of the LF is insensitive to the choice of SF model (within the limited range of models that we have tested here). We will discuss the reasons for this in \S\ref{section:discussion}.

We show our results alongside with a compilation of UV LF constraints on the bright end presented in \citet{Finkelstein2016}, which consist of both ground- and space-based observations from \citet{McLure2009, Castellano2010, VanderBurg2010, McLure2013, Oesch2013, Oesch2014, Schenker2013, Tilvi2013, Bowler2014, Bowler2015, Bouwens2015, Bouwens2016, Finkelstein2015, Schmidt2014, McLeod2015, McLeod2016}. Note that the observational UVLFs are potentially impacted by attenuation by dust in these high-z galaxies, but dust attenuation is not included in the \emph{intrinsic} model UV LF predictions shown in fig. \ref{fig:UVLF_intr}. We notice that the resultant LFs from the GK-Big1 model seem to fit observations fairly well while others overpredict the population of galaxies at $z \lesssim 6$ in the \emph{absence} of dust.  At higher redshifts $z \gtrsim 9$ the results from the KS and GK-Big2 models seem to in better agreement with observations.

\begin{table}
	\centering
	\caption{The best-fit Schechter parameters for the UV LFs inclusive of dust attenuation from our fiducial model between $z = 4 - 10$.}
	\label{table:LF}
	\begin{tabular}{| c | c | c | c | } 
		\hline
		$z$ & $\phi^*$ [$10^{-3}$ Mpc$^{-3}$] & $M^*$ [AB Mag] & $\alpha$   \\
		\hline
		4 & 3.151 & -20.717 & -1.525 \\
		5 & 2.075 & -20.774 & -1.602 \\
		6 & 1.352 & -20.702 & -1.672 \\
		7 & 0.818 & -20.609 & -1.715 \\
		8 & 0.306 & -20.660 & -1.825 \\
		9 & 0.133 & -20.584 & -1.879 \\
		10 & 0.053 & -20.373 & -1.967 \\
		\hline
	\end{tabular}
\end{table}

In fig. \ref{fig:UVLF}, we illustrate the effects of dust attenuation and photoionization squelching by comparing the outputs from our fiducial models with these effects switched on and off. The free parameter $\tau_\text{dust,0}$ in our dust recipe is calibrated to match available constraints from \textit{HST} observations by \citet{Bouwens2014} and \citet{Finkelstein2015}. These observations along with constraints on the faint end from the Frontier Field by \citet{Livermore2017} are presented here for comparison. The dust model predicts that the effect of dust attenuation should be much stronger in bright, massive galaxies given their higher fraction of cold gas and the higher metallicity therein, whilst the effect on faint galaxies is minuscule. We fitted the UV LFs predicted by our fiducial model, with attenuation by dust, with the Schechter function using the least-squares method, where the Schechter function is given by
\begin{equation}
\phi(M)= 0.4\ln(10)\phi^* 10^{-0.4(M-M^*)(\alpha+1)} e^{-10^{-0.4(M-M^*)}}
\label{eqn:schechter}
\end{equation}
\citep{Schechter1976}. The best-fit Schechter parameters are presented in Table \ref{table:LF}.  
\begin{figure}
	\includegraphics[width=\columnwidth]{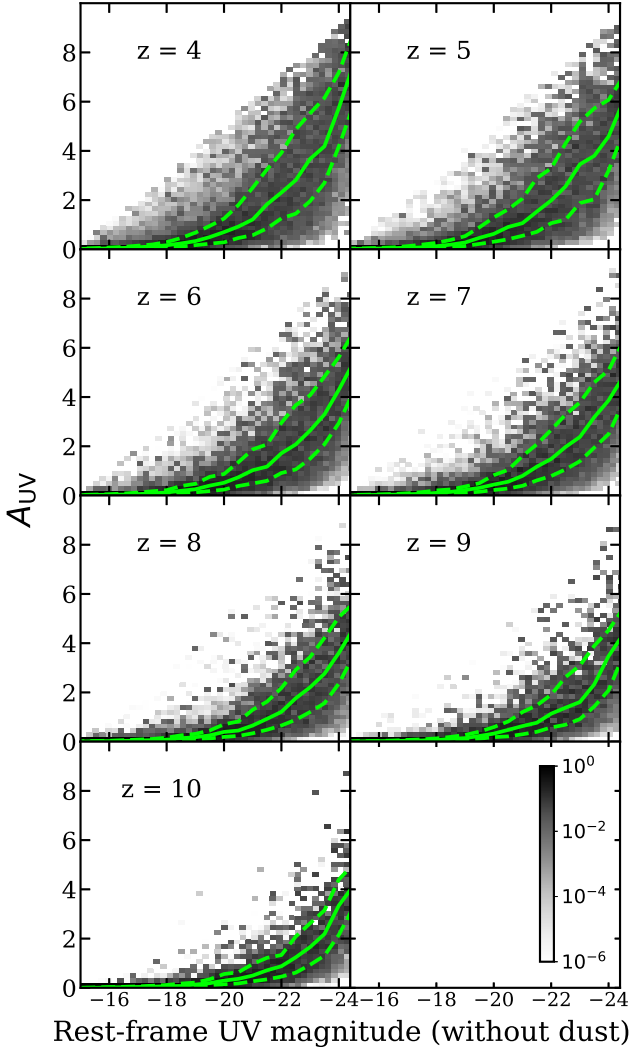}
	\caption{Conditional distributions of extinction in the UV-band versus intrinsic rest-frame UV magnitude in our fiducial model between $z=4$ -- 10. The green solid and dashed lines show the 50th, 16th, and 84th percentiles. The two-dimensional histograms are color-coded to show the conditional number density of galaxies in each bin, which is normalized to the sum of the number density (Mpc$^{-3}$) in its corresponding (vertical) rest-frame UV magnitude bin.}
	\label{fig:luminosity_dust} 
\end{figure}

\begin{figure}
	\includegraphics[width=\columnwidth]{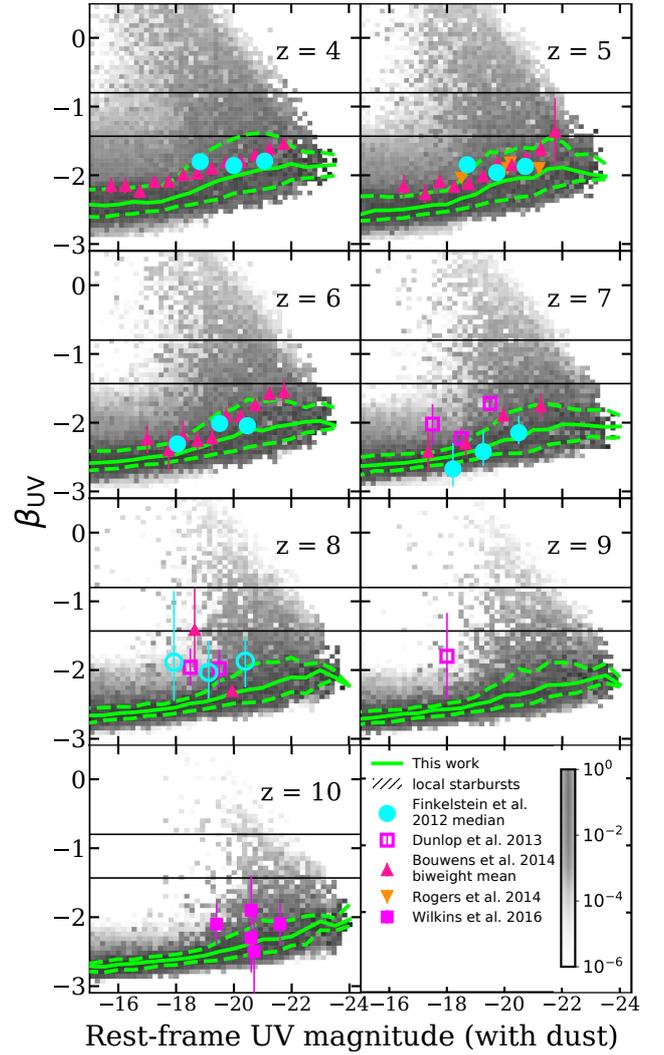}
	\caption{Conditional distributions of the rest-frame UV spectral slope $\beta_\text{UV}$ versus rest-frame UV magnitude (with dust attenuation) in our fiducial model between $z=4$ -- 10. The green solid and dashed lines show the 50th, 16th, and 84th percentiles. The two-dimensional histograms are color-coded to show the conditional number density of galaxies in each bin, which is normalized to the sum of the number density (Mpc$^{-3}$) in its corresponding (vertical) rest-frame UV magnitude bin. Data points are $\beta_\text{UV}$ measurements reported by \citet[cyan circle]{Finkelstein2012}, \citet[open pink square]{Dunlop2013}, \citet[red triangle]{Bouwens2014a}, \citet[yellow square]{Rogers2014}, and \citet[pink square]{Wilkins2016}. The hatched gray band marks where typical local starburst galaxies lie \citep[see text for details]{Finkelstein2012}.}
	\label{fig:beta_dust} 
\end{figure}

In both cases, where dust attenuation is or is not included, our results show that the demographics of luminous galaxies seem to have evolved more rapidly than their faint counterparts. For instance, the ``knee'' feature in the LFs is rather modest at high redshifts but this feature quickly develops and becomes distinct at $z \sim 7$. Moreover, we show that the position of the knee evolves as a function of redshift due to the rapidly evolving dust content. The extinction in the UV-band due to the presence of dust is estimated based on the physical properties of individual galaxies. The UV extinction as a function of $M_\text{UV}$ is illustrated in fig. \ref{fig:luminosity_dust}. The two-dimensional histograms are color-coded according to the conditional number density (Mpc$^{-3}$) of galaxies in each bin, which is normalized to the sum of the number density in its corresponding (vertical) rest-frame UV magnitude bin. The 50th, 16th, and 84th percentiles are marked in each panel to illustrate the statistical distribution. We have verified that the scatter in $A_\text{UV}$ is dominated by the scatter in physical properties that are used to calculate $\tau_{V,0}$ (see Eqn. \ref{eqn:opticaldepth}), whilst the randomly assigned inclination $i$ is sub-dominant. Our results show that even with our simple approach to modeling dust, the scatter can be quite large.

Fig. \ref{fig:beta_dust} shows the $\beta_\text{UV}$-$M_\text{UV}$ relation, where the rest-frame UV luminosity presented here includes dust attenuation, and the photometric rest-frame UV spectral slope $\beta_\text{UV}$ is calculated using the following expression 
\begin{equation}
\begin{split}
\beta_\text{UV} &= \frac{\log(f_{\lambda,\text{FUV}} / f_{\lambda,\text{NUV}})}{\log(\lambda_\text{FUV} / \lambda_\text{NUV})}\\
	&= -0.4 \frac{(m_\text{FUV}-m_\text{NUV})}{\log(\lambda_\text{FUV} / \lambda_\text{NUV})} - 2
\end{split}
\end{equation}
from \citet{Onodera2016}, where $\lambda_\text{FUV}$ and $\lambda_\text{NUV}$ are the central wavelengths of the far- and near-UV bands from the \textit{Galaxy Evolution Explorer} (\textit{GALEX}) survey, where we adopted $\lambda_\text{FUV} \simeq 1530$ \angstrom\ and $\lambda_\text{NUV} \simeq 2300$ \angstrom\ for our calculations. Our results are compared to the range of $\beta_\text{UV}$ spanned by typical local starburst galaxies and the median values of $\beta_\text{UV}$ from a compilation of observations, including GOODS-S DEEP, GOODS-S WIDE, HUDF09, and WFC3 Early Release Science \citep{Koekemoer2011, Grogin2011, Bouwens2010, Oesch2010, Windhorst2011}. We also compare our results to measurements reported by \citet{Dunlop2013} at $z = 7$ -- 9, \citet{Bouwens2014a} at $z = 4$ -- 8, \citet{Rogers2014} at $z = 5$, and \citet{Wilkins2016} at $z = 10$. The predicted $\beta_\text{UV}$-$M_\text{UV}$ relation is is surprisingly good agreement with the observations.  

\begin{figure}
	\includegraphics[width=\columnwidth]{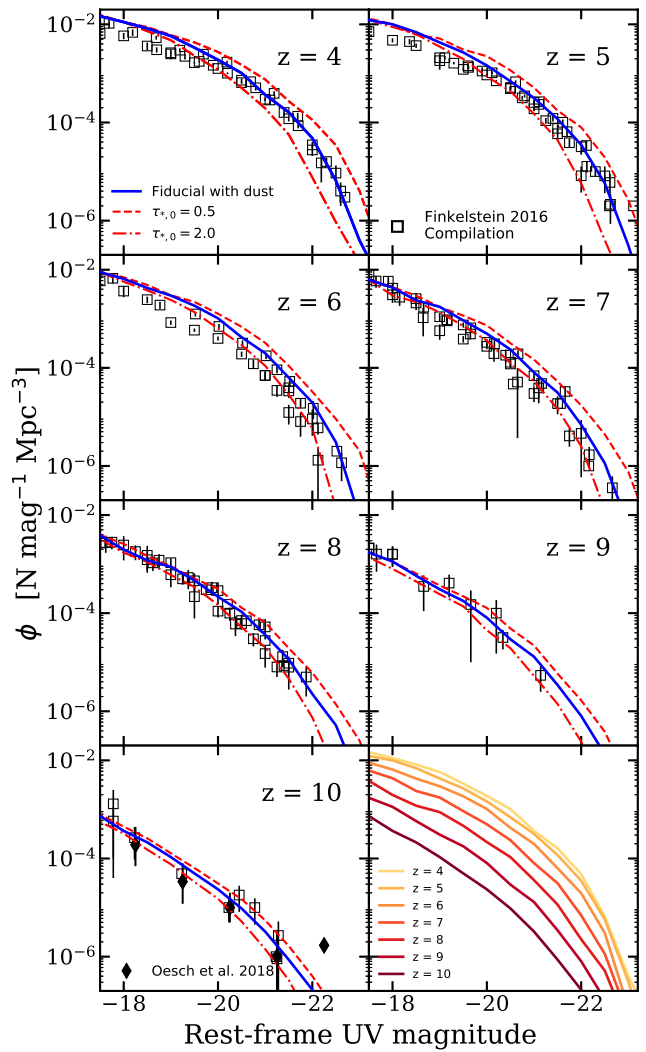}
	\caption{Redshift evolution of the bright end of the UV LFs between $z = 4 - 10$ predicted by our fiducial model with dust attenuation included (blue solid line). Black square markers represent a compilation of observational estimates from space- and ground-based surveys presented in \citet{Finkelstein2016}. Black diamond markers show the additional constraints at $z = 10$ from \citet{Oesch2018}. Red lines represent the cases where we increase or decrease $\tau_{*,0}$ by a factor of 2; dashed and dot-dashed lines are $\tau_{*,0} = 0.5$ and $\tau_{*,0} = 2.0$, respectively. The last panel summarizes the evolution of the bright end of the dust-attenuated UV LFs predicted by the fiducial model.}
	\label{fig:UVLF_bright} 
\end{figure}

In fig. \ref{fig:UVLF_bright}, we zoom into the bright end of the UV LFs and explore the sensitivity of the bright end behavior to the SF timescale $\tau_{*,0}$, where larger $\tau_{*,0}$ means less efficient star formation, and vice versa. This parameter also effectively multiplies the gas consumption timescale, and therefore larger $\tau_{*,0}$ would result in higher gas fractions (again, see \citealt{White2015a}). Here we increase or decrease $\tau_{*,0}$ from our fiducial value by a factor of 2 and find that this results in galaxy populations with $M_\text{UV} \gtrsim -18$ mildly deviating from our fiducial model, with the strength of the deviation seeming to scale with luminosity. Joint constraints on the gas and dust content and SFR in high redshift galaxies from ALMA and JWST will be extremely valuable for breaking the degeneracy between dust and star formation efficiency. With regard to the bright end, we also note that although AGN feedback is responsible for shaping the bright end of the galaxy luminosity function in our models at $z\lesssim 2$, we have checked that switching the AGN feedback on and off has no noticeable effect on any of the results presented here.

\subsection{The faint galaxy populations}
\begin{figure}
	\includegraphics[width=\columnwidth]{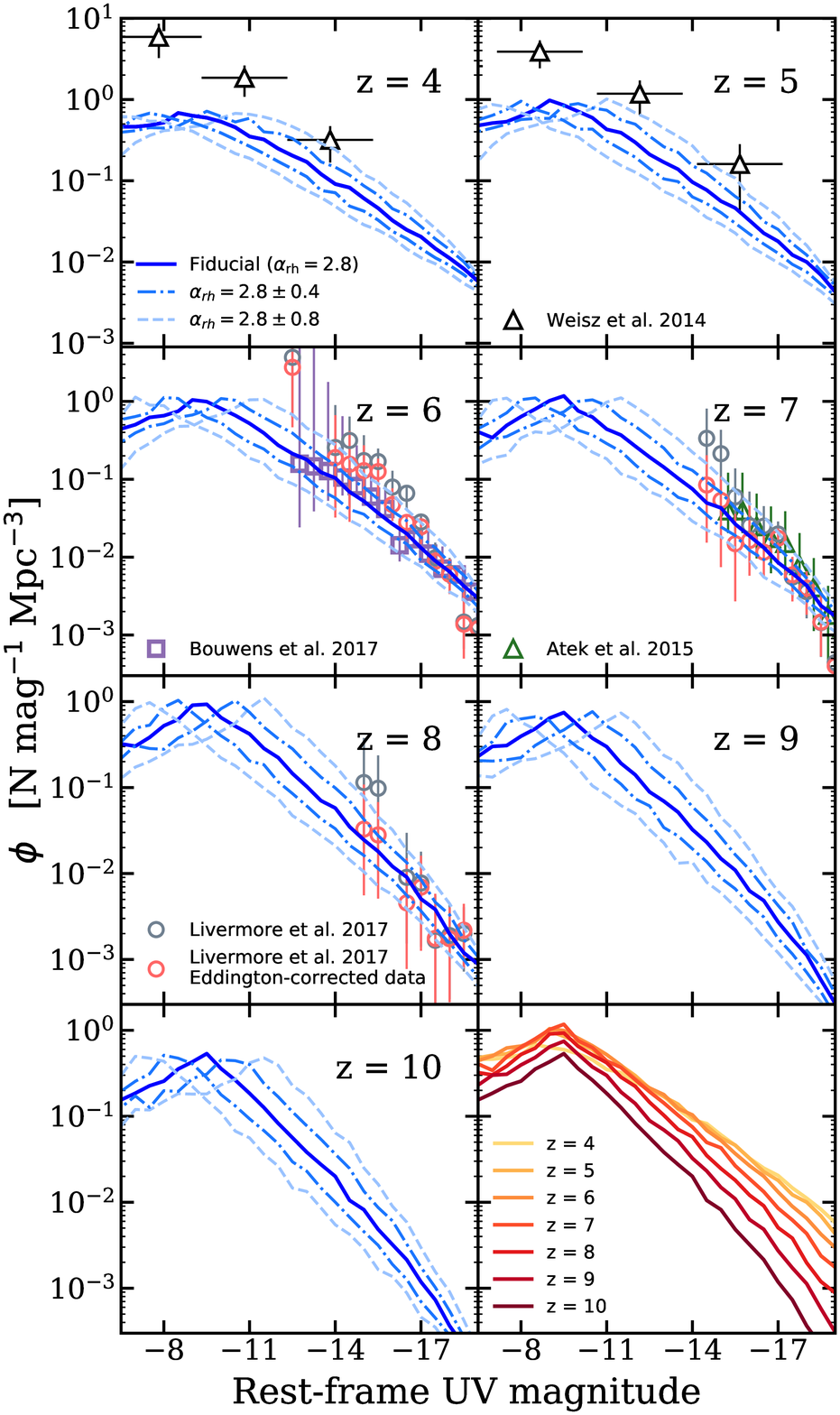}
	\caption{Redshift evolution of the faint end of the UV LFs between $z = 4 - 10$ predicted by our fiducial model with dust attenuation included (blue solid lines). Markers represent the observational estimates from \citet{Livermore2017} as originally published (grey circles) and with Eddington correction (red circles), as well as an independent analysis by \citet[purple squares]{Bouwens2017} for $z\sim6$. We also include estimates from \citet[black triangle]{Weisz2014} and for \citet[green triangle]{Atek2015} for $z\sim7$ (see text for details). We show four additional scenarios where we vary the parameter controlling the mass-loading of stellar driven winds, $\alpha_\text{rh} = 2.8\pm0.4$ and $\pm0.8$. Blue dot-dashed lines show the cases where we let $\alpha_\text{rh} = 2.4$ (above) and $3.2$ (below), and light blue dashed lines show the cases where we let $\alpha_\text{rh} = 2.0$ (above) and $3.6$ (below). The last panel summarizes the evolution of the faint end of the dust-attenuated UV LFs predicted by the fiducial model.}
	\label{fig:UVLF_faint} 
\end{figure}

\begin{figure}
\includegraphics[width=\columnwidth]{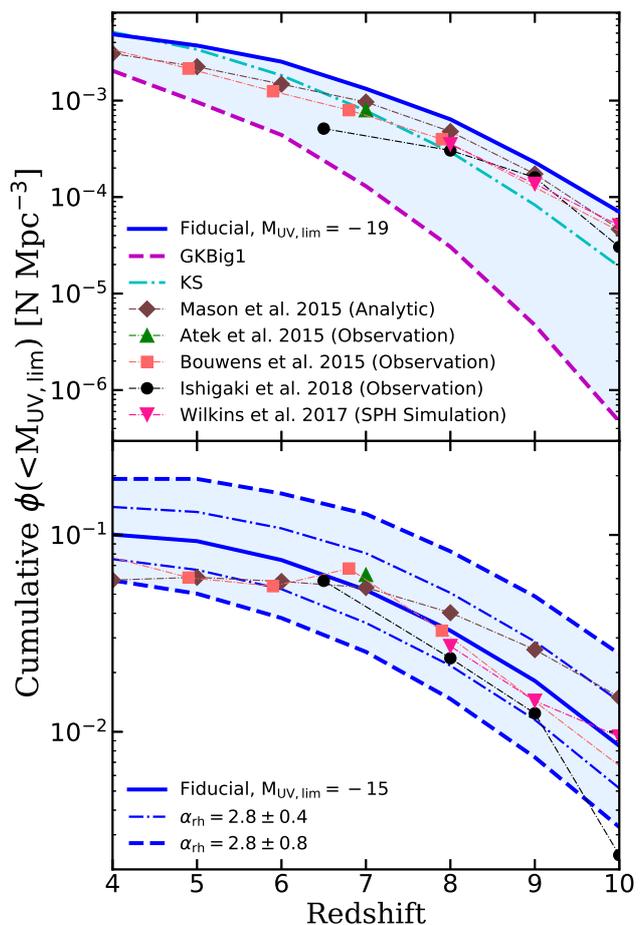}
\caption{The cumulative number density for objects brighter than the specified rest-frame UV luminosities, $M_\text{UV,lim} < -19$. The \textit{upper panel} focuses on the evolution on the bright end. The blue and purple lines show the fiducial and GK-Big1 models, or our most and least optimistic scenarios, respectively. The dashed cyan line shows the result from the KS model. The \textit{lower panel} focuses on faint objects with $M_\text{UV,lim} < -15$. The blue solid line shows the fiducial model. The dot-dashed and dashed lines show the cases where we let $\alpha_\text{rh} = 2.4$ and 2.0 (above), and $\alpha_\text{rh} = 3.2$ and 3.6 (below), respectively. In both panels, the results are compared to a compilation of UV LFs from an SPH simulation, observations, and empirical extrapolations \citep{Mason2015, Atek2015, Bouwens2015, Wilkins2017, Ishigaki2018}.}
\label{fig:cumulative_lf} 
\end{figure}

\begin{figure*}
\includegraphics[width=1.7\columnwidth]{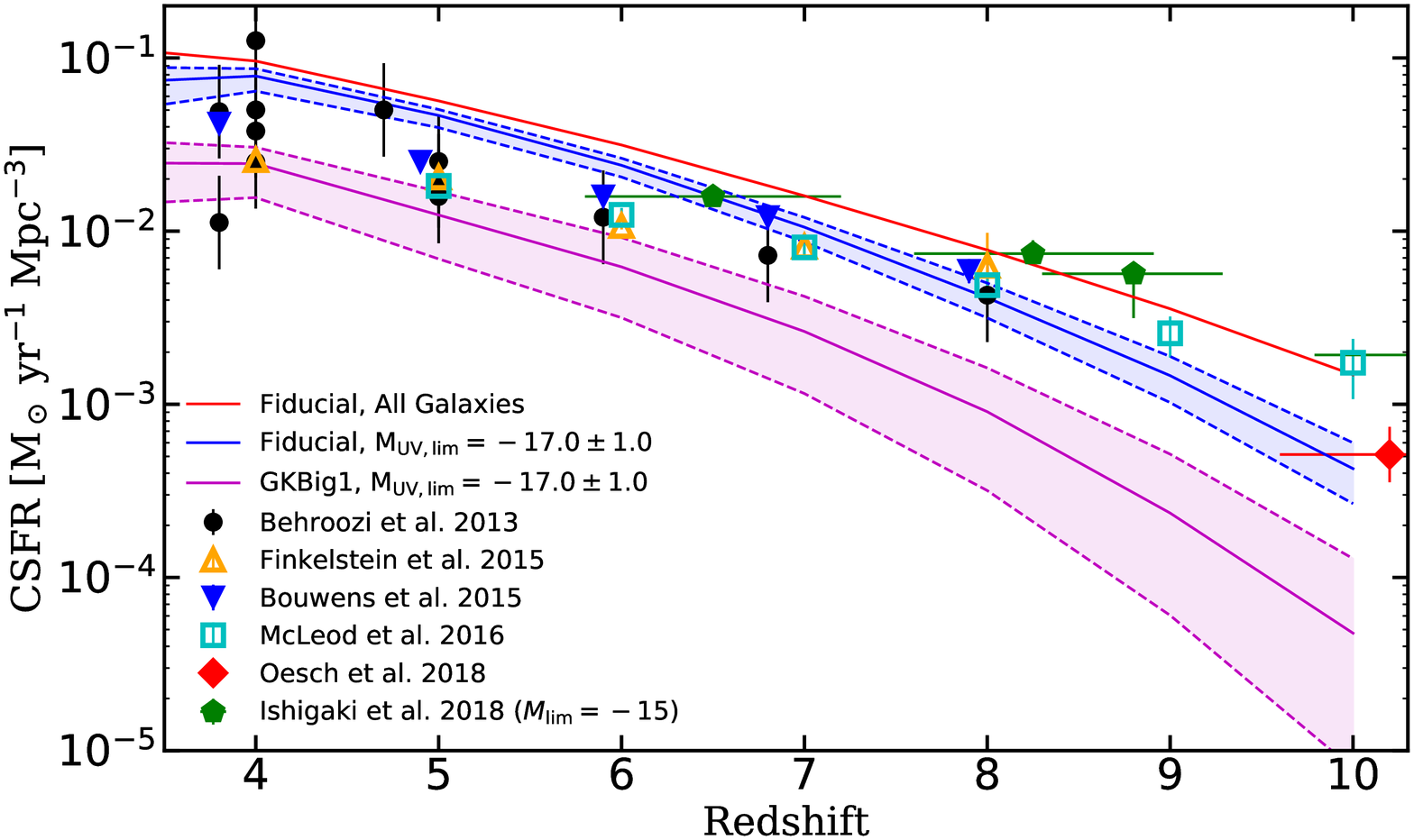}
\caption{Cosmic star formation rate density evolution with redshift integrated down to $M_\text{UV,lim} = -17.0\pm1.0$. The blue and purple lines show results from our fiducial and GKBig1 models, respectively. Our results are compared to observations from other studies by \citet[black circles]{Behroozi2013a}, \citet[orange open triangles]{Finkelstein2015}, \citet[blue inverted triangles]{Bouwens2015}, \citet[cyan open squares]{McLeod2016}, \citet[red diamonds]{Oesch2018}, \citet[green pentagons]{Ishigaki2018}. The cut-off magnitude for these observations is $M_\text{UV} = -17$, unless specified otherwise.}
\label{fig:csfr} 
\end{figure*}

Fig. \ref{fig:UVLF_faint} shows a zoomed-in figure that focuses on the faint end of the UV LFs. Note that the effect of dust is predicted to be negligible in this regime due to the generally low metallicity in these galaxies. The turnover in the LF near $M_\text{UV}\sim-9$ is not due to resolution, but corresponds to the sharp cutoff in the atomic cooling function at 10$^4$ K (which corresponds to $V_{\rm vir} \simeq 17$ \kms).  The slope of the UV LFs remains fairly steep down to this limit. In addition to the estimates of the observed LF from the lensed Frontier Fields, we show constraints on the LF at $z = 4 - 7$ from multiple studies. \citet{Weisz2014} estimate the evolution of the faint end of UV LFs by measuring the star formation histories of 37 Local Group galaxies out to $z\sim5$ using ``fossil evidence'' from resolved stellar populations. Interestingly, their estimates are a factor of $\sim10$ higher than our predictions. The "fossil" method provides a very interesting complementary approach for estimating the faint end of the high redshift luminosity function, but currently it involves some rather uncertain corrections (D. Weisz, private communication). The largest uncertainties likely arise from two factors. The first is the volumetric correction, which attempts to correct for the fact that star formation histories were not available for all known Local Group galaxies, and for Local Group galaxies that are not currently detected. The second issue is that deep resolved color magnitude diagrams (that reach the oldest main sequence turnoff) were not at the time available for many Local Group galaxies. This can lead to a bias that yields increased SF estimates at early times (D. Weisz, priv. comm.). Establishing better links between high-redshift observations and those derived from Local Universe galaxies with resolved stellar population studies is an exciting ongoing area of research, which JWST will also help to advance.

At $z\sim6$ -- 8, we show additional studies based on the Frontier Fields. Here we show both the published data from \citet{Livermore2017} and the unpublished, Eddington bias-corrected number densities for comparison (R. Livermore, priv. comm.). \citet{Bouwens2017} provide constraints on $z\sim6$ galaxies by reanalyzing the Frontier Field observations with a more comprehensive treatment for the magnification and systematic uncertainties. \citet{Atek2015} estimated constraints for LFs using a combined analysis of three lensed fields with their associated parallel fields.

In addition to the fiducial model parameters, we explored the sensitivity of the faint-end slope to the efficiency of stellar driven winds by comparing four additional cases of SN feedback slope $\alpha_\text{rh} = 2.0$, $2.4$, $3.2$, and $3.6$, where larger values of $\alpha_\text{rh}$ imply a steeper dependence of the mass loading factor $\eta$ on galaxy circular velocity, so more gas is ejected from low-mass galaxies. Although these values of $\alpha_\text{rh}$ are not consistent with observations at $z \sim 0$, we present these results as an attempt to quantify the effects of strong stellar feedback on the low-mass galaxy populations and to illustrate how our predictions would change if the effective scalings vary with cosmic time. The results show that the faint end is indeed sensitive to the velocity scaling of the mass loading factor. Higher $\alpha_\text{rh}$ leads to stronger suppression of star formation in low-mass halos, and hence to a flattening of the luminosity function. The adopted value of $\alpha_\text{rh}$ also shifts the luminosity where the LF ``turns over'' from $\sim -8$ (for the ``strongest'') feedback to $\sim -11$ (for the weakest), because $\alpha_\text{rh}$ also affects the slope of the relationship between luminosity (or stellar mass) and halo mass (so the luminosity of galaxies forming in halos with mass corresponding to the atomic cooling limit is shifted up or down). Note that varying the parameter $\epsilon_\text{SN}$ simply shifts the normalization of the whole luminosity function below the knee up or down, as shown in \citet{White2015a}.

Fig. \ref{fig:cumulative_lf} shows the evolution of the cumulative number density of galaxies above some rest-frame luminosity threshold $M_\text{UV,lim}$. First we focus on the evolution of the bright end of the UV LFs, and present the results calculated with $M_\text{UV,lim} = -19$. We show the bracketing cases from our fiducial and GK-Big1 models, which provide the most and least optimistic scenario regarding forming luminous galaxies at high redshift. We compare our results to a compilation of studies from observations, empirical extrapolations, and numerical hydrodynamic simulations \citep{Mason2015, Atek2015, Bouwens2015, Ishigaki2018, Wilkins2017}. This comparison shows that results from other studies are more or less bracketed by our two star formation scenarios.

Additionally, we probe the faint end of the LFs by showing the cumulative number density for galaxies with $M_\text{UV} < -15.0$. Aside from our fiducial model, we include two bracketing cases where we let $\alpha_\text{rh} = $ 2.0 and 3.6, which correspond to weaker and stronger stellar feedback efficiency, respectively. This kind of test will place constraints on the efficiency of stellar winds and whether there are additional dependencies in the wind mass loading scalings beyond the simple assumptions adopted here. The comparison shows that the prediction from our fiducial model is similar to other studies. However, some studies predict fewer galaxies than we predicted at lower redshifts due to a shallower faint-end slope of the UV LFs, which in our model framework would require stronger stellar feedback. 

In fig. \ref{fig:csfr}, we show the cosmic star formation rate (CSFR) for our fiducial and GK-Big1 models, with star formation rate integrated down to $M_\text{UV} = -17 \pm 1.0$. Our result is compared to \citet{Behroozi2013a}, \citet{Finkelstein2015}, \citet{Bouwens2015}, \citet{Bouwens2016}, \citet{McLeod2016}, and \citet{Ishigaki2018}. The CSFR predicted by the KS model is very similar to the ones from our fiducial model, as the UV LF results hinted, and we therefore omitted the KS model from this figure.

Overall, it is intriguing that our models, which were previously calibrated and tested only at much lower redshifts, agree so well with the existing observations all the way out to $z\sim 10$.  Our models predict that the number density of galaxies evolves quite rapidly between $z = 4 - 10$, and that the bright end evolves more rapidly than the faint end. Our models predict that the slope of the UV LFs will remain fairly steep below the current detection limit until $M_\text{UV}\sim-9$. 
This slope is not sensitive to the choice of star formation model but is very sensitive to the scaling of the mass loading factor of stellar-driven winds with galaxy circular velocity, as is the luminosity where the LF turns over at the faint end. The bright end is sensitive to the star formation efficiency or gas depletion timescale and its dependence on gas surface density. However, the effects of dust and the SF efficiency on the bright end of the LF are degenerate, so independent probes of these quantities are needed.

\subsection{Comparison with other models}

In addition to comparing our results with observational constraints, we also compare our predictions with a collection of theoretical studies, including empirical models, semi-analytic models, and cosmological hydrodynamic simulations. Note that studies that are fully or partially numerical are subject to the inevitable tension between simulated volume and spatial and time resolution. For example, simulations with high resolution are only feasible to run over small volumes and rare, massive objects are not well sampled. On the other hand, simulations with coarser resolution and larger volumes do well at capturing formation of large-scale structure. However, reliable predictions for small objects that fall near or below the resolution limit are not possible. For this reason, the galaxy mass and luminosity range covered varies among these studies. And note that among all models compared here, our model has the widest dynamic range and is capable of carrying the simulation to $z = 0$ for comparison with local observations.

Overall, our results agree extremely well with other studies in regimes where observational constraints are available. On the other hand, our predictions for faint galaxy populations, for which observational constraints are very limited due to the lack of direct detections, are in close proximity with a number of other studies, so long as the simulations have adequate resolution to resolve these objects. The faint-end slopes predicted by our fiducial model tend toward the shallower end, especially at higher redshifts, among the range collectively predicted by these models.

Both empirical models included in this comparison employ some empirical relations between halo mass and star formation efficiency that are calibrated using halo abundance matching at redshifts where observational constraints are available. These relations are then extrapolated to make predictions at higher redshifts. The \citeauthor{Mason2015} model, which is based on models introduced by \citet{Trenti2010, Trenti2015, Tacchella2013}, uses a redshift independent star formation efficiency that depends on the halo mass and assembly time, calibrated at $z \sim 5$. The \citeauthor{Sun2016} study explores the outcome of various star formation efficiency models, including one that assumes a power-law extrapolation below a cutoff halo mass limit $M_\text{h} = 2 \times 10^{10}$ \Msun\ and a best-fit redshift independent model. Note that their models are calibrated to bright galaxies at $z = 6$ -- 8,  and results at $z = 9$ and 10 are predictions that are yet to be published. 

On the other hand, the details for the numerical hydrodynamic simulations included in this comparison are summarized in the following. \textsc{BlueTides} is a large-volume cosmological hydrodynamic simulation that focuses on the high-redshift ($z \gtrsim 8$) universe, with a box that is 400 Mpc $h^{-1}$ on a side, resolving galaxies with $M_* \gtrsim 10^8$ \Msun\ toward the end of their simulation \citep{Wilkins2017}. \textsc{Vulcan} is a high-resolution (dark matter particle $\sim 10^5$ \Msun), relatively small volume (25 comoving Mpc on a side) simulation that aims to quantify the contribution of faint galaxies to cosmic reionization \citep{Anderson2017}. The Cosmic Dawn (CoDa) simulation is a large-scale hydrodynamic simulation coupled with radiative transfer modeling, resolving galaxies down to $\sim 10^8$ \Msun\ in box that is $\sim 100$ Mpc on a side \citep{Ocvirk2016}. Cosmic Reionization On Computers (CROC) is a cosmological simulation based on the adaptive refinement tree (ART) method \citep{Gnedin2014, Gnedin2014a}. The CROC results compared here are from their fiducial simulation box with 20 Mpc $h^{-1}$ on a side and 512$^3$ initial grid cells, with spacial resolution of 100 pc \citep[see][]{Gnedin2016}. The DRAGONS project \citep{Poole2016} consists of a SAM \textsc{meraxes} that is built on top of the \textit{Tiamat} suite of $n$-body simulations. We also include results from a high-resolution hydrodynamic simulation with embedded self-consistent radiative transfer that is similar to the one described in \citet{Finlator2012,Finlator2015,Finlator2016,Finlator2017}. The three runs included here have box sizes of 6.0, 7.5, and 10.0 Mpc $h^{-1}$ on a side, with $2\times256^3$, $2\times320^3$, and $2\times512^3$ particles; the UV background is discretized into 32, 40, and 64 voxels. 

\begin{figure}
	\includegraphics[width=\columnwidth]{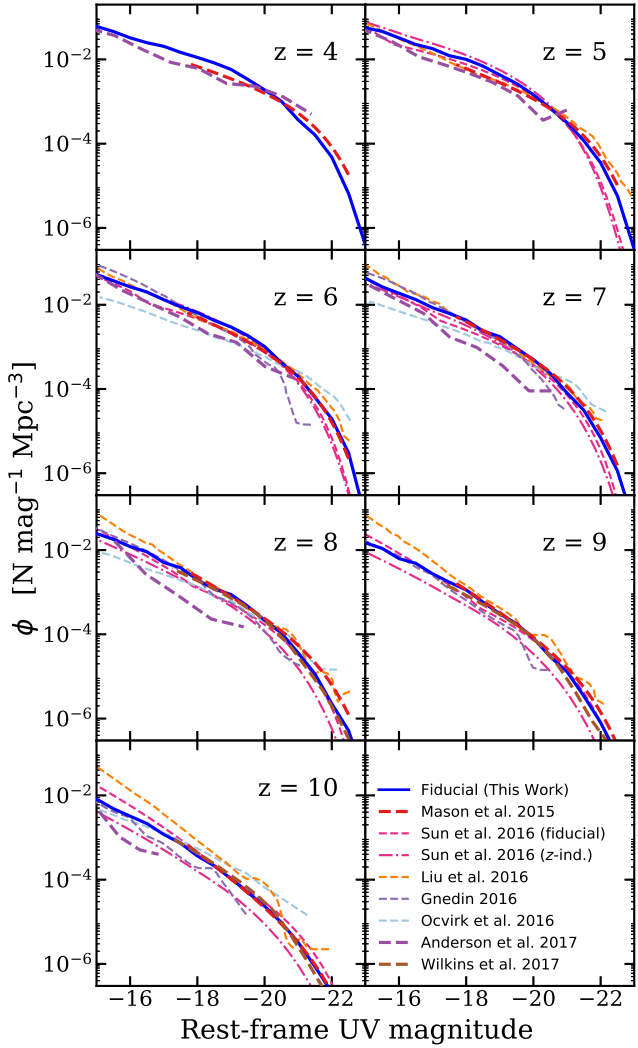}
	\caption{Redshift evolution of the bright end of the UV LFs between $z=4$ -- 10 predicted by our fiducial model with dust attenuation (blue solid lines). Our results are compared to several other theoretical studies, including empirical studies from \citet[green]{Mason2015} at $z=4$ -- 10 and \citet[red]{Sun2016} at $z=5$ -- 10 from their fiducial model (dashed) and redshift-independent model (dot-dashed), cosmological hydrodynamic simulations \textsc{BlueTides} \citep[brown]{Wilkins2017} at $z=8$ -- 10 and \textsc{Vulcan} \citep[purple]{Anderson2017} at $z=4$ -- 10, and the DRAGONS SAM simulation suite \citep[orange]{Liu2016} at $z=5$ -- 10. See text for details.}
	\label{fig:UVLF_models_bright} 
\end{figure}

\begin{figure}
	\includegraphics[width=\columnwidth]{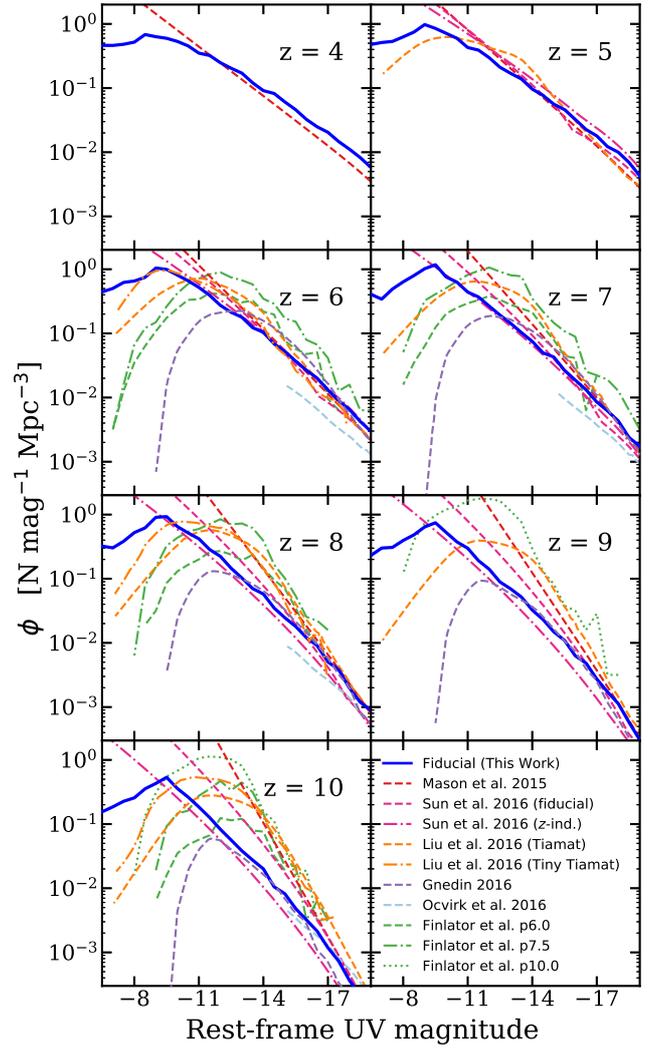}
	\caption{Redshift evolution of the faint end of the UV LFs between $z=4$ -- 10 predicted by our fiducial model with dust attenuation (blue solid line). Our results are compared to several other studies, including empirical models from \citet[red]{Sun2016} at $z=5$ -- 10 fiducial model (dashed) and $z$-independent model (dot-dashed), DRAGONS based on their \textit{Tiamat} (dashed) and \textit{Tiny Tiamat} (dot-dashed) simulations at $z=5$ -- 10 \citep[orange]{Liu2016}.  We also include results from a hydrodynamic simulation with embedded self-consistent radiative transfer that is similar to the one presented in \citet[green]{Finlator2012} at $z=6$ -- 10 in three different box sizes and resolutions (dashed, dot-dashed, and dotted in increasing resolution). See text for more details.}
	\label{fig:UVLF_models_faint} 
\end{figure}

In fig. \ref{fig:UVLF_models_bright}, we compare the bright end of our UV LFs to the predictions of \citeauthor{Mason2015}, \citeauthor{Sun2016}, DRAGONS, CROC, CoDa, \textsc{Vulcan}, and \textsc{BlueTides}. In fig. \ref{fig:UVLF_models_faint}, we compare our predictions for the faint-end behavior of the UV LFs  to \citeauthor{Mason2015}, \citeauthor{Sun2016}, DRAGONS, CROC, and \citeauthor{Finlator2012}. The \citeauthor{Mason2015} model assumed a $M_\text{UV} = -12$ cutoff for the atomic cooling limit in their work and the \citeauthor{Sun2016} model considered a limiting magnitude of $M_\text{UV} = -13$ in their CSFR calculation. For this comparison, we plot the results from these analytic studies extrapolated down to fainter magnitudes to illustrate the faint-end slopes predicted by these models and how they stack up with other numerical results. On the other hand, numerical simulations provide predictions with physical processes traced self-consistently as far down as resolution permits, although they are still limited by the assumptions inherent in their sub-resolution recipes for star formation and stellar feedback. Moreover, we note that the very faint populations predicted by these numerical studies might be somewhat resolution dependent, rather than physical. For instance, some subtle differences can be spotted between the UV LFs based on the coarser \textit{Tiamat} (at $z = 6$ -- 10) and the finer \textit{Tiny Tiamat} (at $z = 6$, 8, and 10) $N$-body simulations from the DRAGONS simulation suite in our comparison \citep{Liu2016}. Similar behavior can be seen among the three different runs in the \citeauthor{Finlator2012} simulations. However, heating from a photoionizing background is tracked self-consistently with on-the-fly radiative transfer in this model. The \textit{Renaissance Simulations} \citep{OShea2015}, SPHYNX \citep{Cabezon2017}, and FIRE \citep{Ma2017} are theoretical studies that address similar issues but are not included in our comparison here.

\section{Predicted Apparent Magnitude Functions with NIRCam Filters}
\label{section:nircam}

\begin{figure}
\includegraphics[width=\columnwidth]{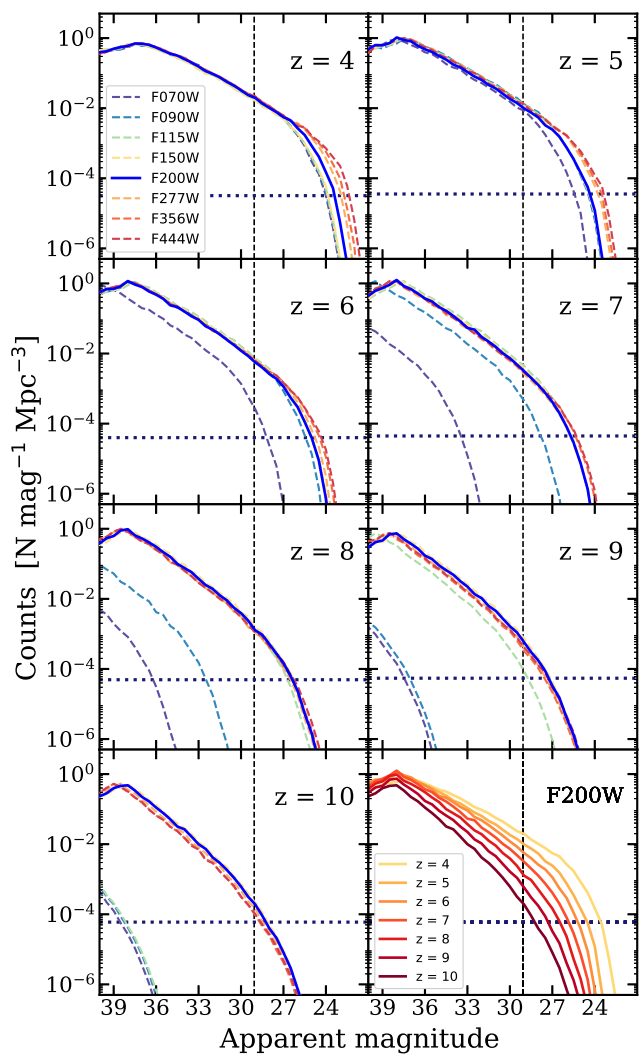}
\caption{Redshift evolution of the AMFs as seen in every NIRCam filter between $z = 4 - 10$. See Table \ref{table:filter_fact} for the specifications for the filters compared here. The vertical black dashed lines represents the detection limit of NIRCam assuming a $10^4$ second exposure. The horizontal dashed line shows where one object is expected per NIRCam field of view ($2\times2.2\times2.2$ arcmin$^2$). The last panel summarizes the evolution of the AMF for our nominal F200W filter.}
\label{fig:AMF} 
\end{figure}

\begin{figure}
\includegraphics[width=\columnwidth]{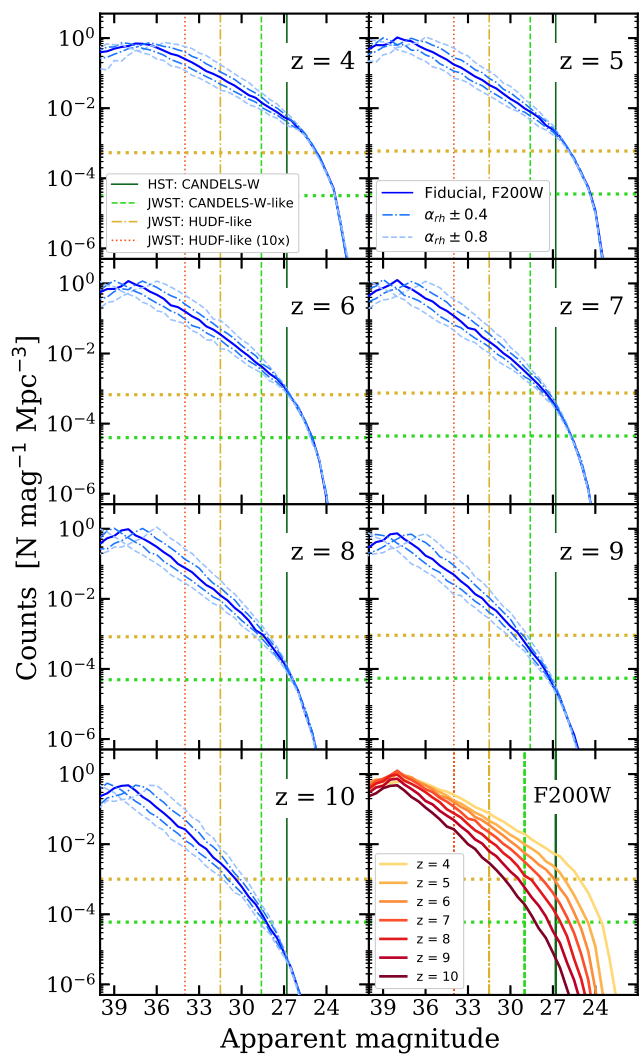}
\caption{Redshift evolution of the AMFs in the F200W filter. The vertical dashed lines represent the detection limits for example \textit{JWST} surveys similar to legacy \textit{HST} counterparts; see Table \ref{table:survey_fact} for details. The green and yellow horizontal dashed lines show where ten objects are expected in a $\sim 100$ arcmin$^2$ survey and in the HUDF field ($2.4\times2.4$ arcmin$^2$), respectively. Blue dashed lines show the cases where we let $\alpha_\text{rh} = 2.4$ (above) and $3.2$ (below), and light blue dot-dashed lines show the cases where we let $\alpha_\text{rh} = 2.0$ (above) and $3.6$ (below). The last panel summarizes the evolution of the AMF for the F200W filter. }
\label{fig:AMF_2} 
\end{figure}

In this section, we provide the distribution functions of apparent magnitude (AMF) for high-redshift galaxy populations, where the latest published \textit{JWST} NIRCam filter response functions are used to compute the apparent magnitude for the galaxies predicted by our model. We also provide estimates of the number of detected objects for several example survey configurations.

\begin{table}
	\centering
	\caption{Pivot wavelengths and bandwidths of NIRCam filters. Detection limits assuming a $10^4$ seconds exposure. $^a$\textit{JWST} User Documentation\protect\footnotemark}
	\label{table:filter_fact}
	\begin{tabular}{ | l | c | c | c | } 
		\hline
		NIRCam & $\lambda$$^a$ & Bandwidth$^a$ & Detection Limit\\
		Filters & [$\mu$m] & [$\mu$m] & [AB Mag] \\
		\hline
		F070W & 0.704 & 0.132 & 28.16 \\ 
		F090W & 0.902 & 0.194 & 28.56 \\ 
		F115W & 1.154 & 0.225 & 28.85 \\
		F150W & 1.501 & 0.318 & 29.04 \\
		F200W & 1.989 & 0.457 & 29.07 \\
		F277W & 2.762 & 0.683 & 28.93 \\
		F356W & 3.568 & 0.781 & 28.93 \\
		F444W & 4.408 & 1.029 & 28.33 \\
		\hline
	\end{tabular}
\end{table}
\footnotetext{https://jwst-docs.stsci.edu/display/JTI/NIRCam+Filters}

In fig. \ref{fig:AMF}, we showcase the evolution of the AMFs over redshift as seen in the eight NIRCam broadband filters: F070W, F090W, F115W, F150W, F200W, F277W, F356W, and F444W. The central wavelength and the detection limit of these filters are summarized in Table \ref{table:filter_fact}. Detection limits are 10-$\sigma$ point source limiting AB magnitudes from the \textit{JWST} Exposure Time Calculator\footnote{\url{https://jwst.etc.stsci.edu/}, v1.2.2} for an exposure time of $10^4$ second, measured in 0.04'' diameter apertures for the short-wavelength camera, and 0.08'' diameter apertures for the long-wavelength camera, assuming a low background level. An approximate detection limit $m_\text{lim} \sim 29$ is marked in the figure with a black dashed line. Objects to the left of the line are too faint to be detected at this depth. We also marked where one object is expected per NIRCam field of view ($2\times2.2\times2.2$ arcmin$^2$) with a horizontal dashed line in each panel. Objects with counts below this line are too rare to be found on average in a single pointing of NIRCam. The seven panels together demonstrate how the high-redshift galaxy populations quickly drop out from the filters with shorter wavelengths due to absorption by the intervening IGM. Our results show that the F200W filter, our choice of nominal filter, is best suited for detecting objects across all redshifts of interest, and might even be able to pick up bright objects beyond $z \sim 10$ with an extended exposure time. The Schechter function fitting parameters for the F200W filter are presented in Table \ref{table:AMF}. Redundant detections from multiple filters also improve the accuracy of estimates of dust attenuation, which is essential to properly uncover the underlying intrinsic UV luminosity. Galaxies at high redshifts rapidly drop out of the shorter wavelength filters F070W, F090W, and F115W for $z < 10$, and F150W at $z \sim 12$, whilst F200W, F277W, F356W, and F444W remain advantageous for possible detections up to $z \sim 15$ given sufficient exposure time. Although MIRI is designed to detect longer wavelength radiation, it will not be able to detect such high redshift galaxies due to its low sensitivity. 

\begin{table}
	\centering
	\caption{Detection limits for selected \textit{HST} blank field surveys and for comparable anticipated \textit{JWST} legacy surveys. \textit{JWST} prediction limits are estimated assuming the use of the F200W filter. At the end we show a bracketing extreme case where \textit{JWST} conducts a HUDF-like survey on a cluster field with 10x magnification. }
	\label{table:survey_fact}
	\begin{tabular}{| l | l | c | c | c | }
		\hline
		& Survey Name & Detection Limit$^{a}$ \\
		& & [AB Mag] \\
		\hline
		\textit{HST} & CANDELS - Wide   & 26.8 \\
		& CANDELS - Deep   & 27.8 \\
		& HUDF   & 29.5 \\
		& Frontier Field (unlensed) & 29.0 \\
		& Frontier Field (10x mag.) & 31.5 \\
		\hline
		\textit{JWST} & CANDELS-Wide-like & 28.6 \\
		& HUDF-like & 31.5 \\ 
		& HUDF-like (10x mag.) & 34.0 \\ 
		\hline
	\end{tabular}
\end{table}

In fig. \ref{fig:AMF_2}, we further investigate the population of galaxies that are expected to be found by \textit{JWST} in different surveys. We show the same AMFs from the F200W filter and provide the detection limits from a number of simulated surveys. The detection limits of past \textit{HST} surveys and upcoming \textit{JWST} surveys are summarized in Table \ref{table:survey_fact}. First, we show the detection limit of the \textit{HST} CANDELS-Wide survey for reference. Then, from bright to faint, we show the detection limits estimated for a \textit{JWST} investment comparable to CANDELS-W and HUDF, and the very optimistic case where we gain a factor of 10$\times$ magnification from gravitational lensing from a massive galaxy cluster. We indicated where $\sim10$ objects are expected to be found in survey areas similar to that of the HUDF field ($2.4\times2.4$arcmin$^{2}$) and CANDELS-W ($\sim100$ arcmin$^{2}$), which are both estimated assuming $dz = 1$ slices centered at the given redshift. We also show the cases where we deviate from the fiducial value and let $\alpha_\text{rh} = 2.8 \pm 0.4$ and $\pm 0.8$, where stronger feedback gives shallower faint-end slopes and vice versa. This comparison shows that \textit{JWST} will be able to reliably constraint the faint-end slope up to $z\sim8$, or even $z\sim10$ in lensed fields. This will provide important constraints on stellar feedback physics.

\begin{table}
	\centering
	\caption{The best-fit Schechter parameters for the AMFs for the NIRCam F200W filter between $z = 4 - 10$ predicted by our fiducial model.}
	\label{table:AMF}
	\begin{tabular}{| c | c | c | c | } 
		\hline
		$z$ & $\phi^*$ [$10^{-3}$ Mpc$^{-3}$] & $M^*$ [AB Mag] & $\alpha$  \\
		\hline
		4 & 2.423 & 24.748 & -1.526 \\
		5 & 1.709 & 25.576 & -1.589 \\
		6 & 0.989 & 25.955 & -1.676 \\
		7 & 0.569 & 26.281 & -1.733 \\
		8 & 0.222 & 26.427 & -1.830 \\
		9 & 0.113 & 26.767 & -1.881 \\
		10 & 0.052 & 27.163 & -1.950 \\
		\hline
	\end{tabular}
\end{table}

\begin{figure}
\includegraphics[width=\columnwidth]{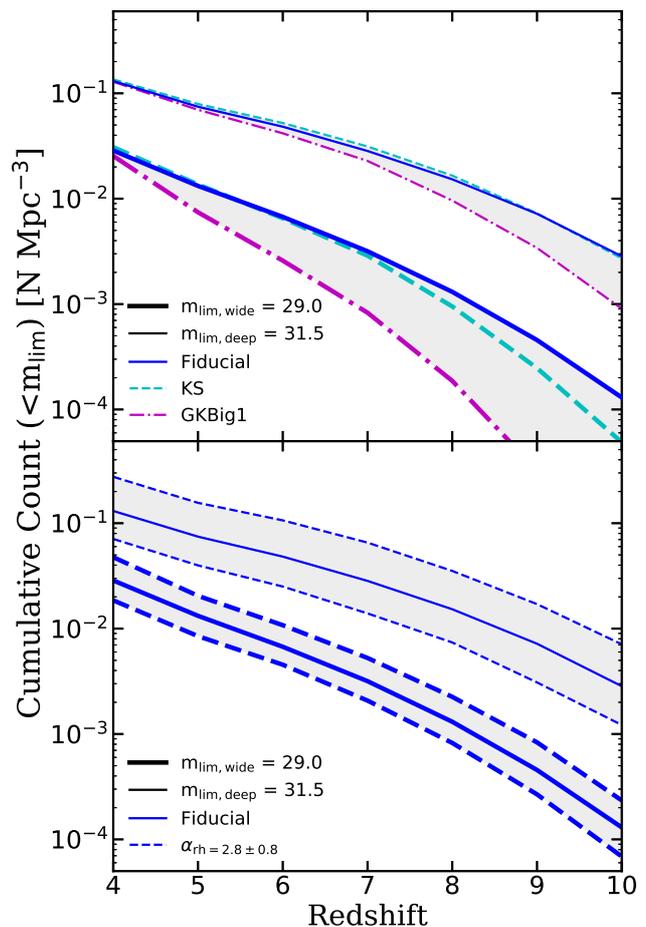}
\caption{The number density of objects brighter than some specified apparent magnitudes in the NIRCam F200W filter, $m_\text{lim}$, where the bold (narrow) line shows $m_\text{lim,wide} = 29.0$ (31.5) that is chosen to be close to the detection limit of a JWST investment similar to CANDELS-W (HUDF). The blue solid and dashed lines in both panels are identical and they serve as a visual guide for the predictions from our fiducial model. In the \emph{upper panel}, the purple dot-dashed line shows the GK-Big1 model and the cyan dashed line shows the KS model. In the \emph{lower panel}, the cyan dashed line shows the results for $\alpha_\text{rh} = 2.0$ and the purple dot-dashed line shows the results for $\alpha_\text{rh} = 3.6$. Light gray bands are added to visually group the lines that share the same $m_\text{lim}$.}
\label{fig:obj_count} 
\end{figure}

In fig.\ref{fig:obj_count}, we show the number of objects per Mpc$^{3}$ above some detection limits, $m_\text{lim}$, expected to be found using the NIRCam F200W filter. We consider two cases where we let $m_\text{lim,wide} = 29$ and $m_\text{lim,deep} = 31.5$, representing the expected detection limit for \textit{JWST} wide-field surveys (e.g. CANDELS-Wide) and deep-field surveys (e.g. HUDF). In the upper panel, we illustrate the difference among the three SF models considered in this work, and in the lower panel we show the cases where we let $\alpha_\text{rh} = 2.0$ and $\alpha_\text{rh} = 3.6$. Here, a more direct comparison with observations can be made, without the need for applying $K$-corrections to the observations to get from observed to rest-frame magnitudes. One can see that, similarly, observations of the bright population will constrain the star formation efficiency, while observations of the faint end will constrain stellar driven winds.

\section{Discussion}
\label{section:discussion}

The formation and evolution of galaxies are governed by a complex network of intertwined physical processes that operate over many physical scales. Although many details remain to be worked out, the community seems to have reached a broad consensus regarding the main physical processes that shape galaxy properties at $z \gtrsim 6$; these include cosmological accretion, strong stellar-driven winds that are more efficient at low masses, and (more controversially), black hole feedback that preferentially suppresses star formation at high masses. However, these processes can sometimes have degenerate effects on galaxy properties. Although differentiating the effects of these processes is extremely challenging, being able to do so is crucial for interpreting galaxy observations at all redshifts. Observational constraints at $z \gtrsim 6$ are currently quite limited, leaving significant uncertainties in the current theories of galaxy formation and their predictions at very high redshifts. One of the main highlights of this work is that we explored how variations in several uncertain physical processes in theoretical models will impact the global physical properties of galaxies and what may be seen by future observations. We also discuss how these processes may be disentangled with the significant insights \textit{JWST} observations have to offer.

High-$z$ star-forming galaxies are most luminous in the observed frame infrared, making them observable with NIRCam and MIRI onboard \textit{JWST}, and it has been shown that the choice of broadband filters can have an impact on the photometric redshift estimation and the follow-up interpretation \citep{Bisigello2016, Bisigello2017}. In fig. \ref{fig:AMF}, we demonstrated how objects drop out from shorter wavelength filters starting at $z \sim 5$, making the F200W filter the nominal choice for blind surveys for these deep-field objects across $z = 4$ -- 10 with investments comparable to CANDELS-Wide and HUDF. For galaxies beyond $z \sim 10$, observations in the longer wavelength filters, namely F356W and F444W, will be required with exposure times $\gtrsim 10^5$ seconds. 

With its unprecedented sensitivity, \textit{JWST} will be able to probe large populations of galaxies in deep space and provide constraints for the faint-end slope of the UV LFs at $z \lesssim 8$ and for the evolution of the bright-end up to $z \sim 10$. These observations will put the empirical relations derived from low-redshift observations to the test in extreme conditions. Moreover, the redundancy in multiple NIRCam broadband filters at lower redshifts will enable multiwavelength measurements, which is crucial to recovering the physical properties of high-$z$ galaxies and breaking degeneracies in the underlying physical processes.

In this work, we utilized a computationally efficient SAM to make quantitative estimates of observable properties for galaxy populations expected to be detectable with \textit{JWST}. The physically motivated empirical recipes adopted in the model are well-tested at lower redshifts ($z \lesssim 6$, see \citetalias{Somerville2015}). We present model predictions at redshifts where these models have never been tested before, and where observational constraints are relatively limited. By exploiting the efficiency of our model, we also provide forecasts for various model variations, including exploring multiple SF recipes and parameterizations of outflow rates for stellar-driven winds.

Physical processes that shape the formation of galaxies are degenerate, and yet each of them evolves slightly differently over time and can simultaneously affect multiple physical properties. Traditionally, there has been tension for galaxy formation models to simultaneously match the observed gas fraction, stellar metallicity, and stellar fraction. In order to calibrate our model, we carefully balance the model parameters for multiple physical processes to match observations at $z \sim 0$. Fig. \ref{fig:calibration} summarizes how the outputs of our calibrated model compare with observational constraints. See appendix for details of the calibration process and tests.

Massive star forming galaxies in the lower redshift Universe are obscured by dust in the line of sight, which makes it extremely difficult to determine the true underlying stellar content. In our model, the attenuation effect due to dust in the galactic disc is estimated based on galaxy physical properties, incorporated with a dust-to-metal ratio guided by observations. In agreement with previous studies, we find that the optical depth of dust, or effectively the dust-to-metal ratio, is required to evolve with redshift. However, we also note that our very simple dust recipe does not accurately represent the complex geometries of dust relative to stars that may be common in high redshift galaxies \citep{Koprowski2016, Chen2015}.

We find that star formation remains fairly efficient in low-mass halos, in conflict with the \citet{Krumholz2012}  model, which argues that star formation is heavily suppressed at $z > 2$ in dark matter halos with masses $< 10^{11}$ \Msun\ due to the low metallicity of gas in these halos. However, recent lensed and very deep fields observations provide strong support for LF remaining fairly steep down to low halo.  masses. We are aware of the fact that the underlying assumptions of these metallicity-based multiphase-gas partitioning recipes tend to break down in extremely low-metallicity environments (e.g. $Z_\text{cold} < 0.05 Z_\odot$ \citep{Krumholz2009, Gnedin2011}, which might cause our model to over-predict the stellar content in our least massive halos and at higher redshifts.

Cosmic reionization in our models is treated with a rather ad-hoc prescription, in which the whole universe is reionized uniformly and instantaneously. Our models currently do not attempt to model the formation of metal-free Population III stars explicitly. Instead, we set a metallicity floor of  $Z_\text{pre-enrich}$ to represent pre-enrichment in the initial hot gas in halos and the gas accreted onto halos due to cosmological infall.

\subsection{Physical processes that shape the bright end of UV LFs}

The strong evolution in the bright end of the UV LFs indicates a rapid growth of massive galaxies in the early universe. However, the physical processes that drive or regulate this evolution are still unclear. Due to the limited direct observations at $z \gtrsim 6$, it is extremely difficult to disentangle the effects of multiple processes.

Each observed property of galaxies is resulting from a unique combination of underlying physical processes. Although the underlying fundamental physics should remain the same at all times, since the physical processes that drive the evolution of these objects depend on a number of effectively redshift dependent conditions, the global properties of galaxies may also be effectively redshift dependent. The evolution of these resulting global physical properties change subtly depending on the strength of the impact of each of the physical processes. Therefore, being able to disentangle the impact of these driving physical processes, as well as the evolution of the resultant observable properties, is most important for understanding galaxy formation and for making accurate predictions for the galaxies that are yet to be directly observed. 

In this work, we have shown that the choice of star formation recipe and the gas depletion time both have a very strong impact on the abundance of bright galaxies. We have also provided model outputs where we systematically vary some of the physical recipes and associated parameters. For instance, changing the gas depletion timescale by a factor of two in either direction seems to mainly impact galaxies of $M_\text{UV} \lesssim  -19$.  Under this variation, the bright end of the UV LFs deviates mildly from the fiducial results but still remains mostly within the uncertainties in the observations. On the other hand, altering the SF recipes would significantly change the predicted number density for luminous galaxy populations, where the distinction occurs at $M_\text{UV} \lesssim -14$ at $z = 10$, and this limit gradually evolves to $M_\text{UV} \lesssim -20$ at $z = 4$. The abundance of faint galaxies is largely insensitive to either of these processes due to self-regulation of SF in low-mass halos, which will be discussed in detail in the next subsection.

To focus on the impact of these processes on the underlying stellar populations, in fig. \ref{fig:UVLF_intr} we compare the intrinsic UV LFs in the absence of dust. The free parameters in the empirical SF recipes that are calibrated to $z \sim 0$ observations and the SF relation do not evolve over time, with the fiducial GK-Big2 and GK-Big1 representing the more and less optimistic SF scenarios, respectively. Both our fiducial model and the classic \citetalias{Kennicutt1998} model are able to match high-$z$ observations quite well without further parameter retuning. Given that the general consensus is that high-$z$ galaxies are much less dusty than their low-$z$ counterparts, these empirical SF recipes might hold to a certain extent even in extreme environments. Interestingly, GK-Big1 is able to match observations quite well at low redshifts ($z \lesssim 6$) when dust attenuation is omitted, but it underpredicts the number of bright galaxies at higher redshifts. This implies that if the dust attenuation at these redshifts is significantly lower than what we have assumed, a mildly evolving $\Sigma_\text{SFR}$--$\Sigma_\text{\molh}$ relation may be able to produce sufficient numbers of bright galaxies at high redshifts.

It is well known that at $z\sim 2$ -- 4, the most rapidly star-forming galaxies are heavily obscured by dust. The dust content of galaxies at higher redshift is uncertain. Since dust is built up over generations of star formation, the general expectation is that high redshift galaxies should be less dusty than lower redshift ones \citep{Popping2017}. Moreover, there is mounting evidence that the dust geometry may be different in high redshift galaxies, giving rise to differences in attenuation for a given dust mass \citep{Popping2017a, Narayanan2018}. Most theoretical models that include the effects of dust assume a relationship between dust optical depth and the metallicity and gas surface density, similar to the one we have adopted here. Our simple empirical model for dust is calibrated guided by available observations at $z = 4$ -- 10, which requires a redshift dependent dust-to-metal ratio in order to be able to simultaneously match constraints at all redshifts within the framework of our simple model. Our results show that dust attenuation seems to be less effective at $M_\text{UV} \gtrsim -19$, however, with a much stronger redshift evolution comparing to bracketing cases for gas depletion times we showed.  Future multiwavelength observations  with multiple NIRCam filters may help constrain the dust content (e.g. using the UV continuum slope $\beta$ as in \citealt{Finkelstein2012}, also see \citealt{Popping2017a}).

Using similar SAMs and a similar dust model, \citetalias{Somerville2012} found that their models with a Calzetti attenuation curve underproduced UV-luminous galaxies at high redshift ($z\sim 3$--5). Our updated models perform better in this regard, as a result of the updated Planck cosmology and associated recalibration. As shown in fig. \ref{fig:tau_dust_0}, the amount of dust required to match observations between $ = 4$ -- 6 has been reduced by roughly a factor of two. However, the UV LFs remain quite sensitive to $\tau_\text{dust,0}$ and, as pointed out in previous works, using a fixed dust parameter that is normalized to observations at low redshifts would systematically underproduce UV-luminous galaxies at high redshift. In future works we plan to explore models with a more self-consistent treatment of dust formation and destruction and more complex treatment of dust geometry, and the implications for Far-IR and mm and sub-mm observations (Popping et al. in prep).

Although we find that black hole feedback plays a negligible role in shaping galaxy properties at these redshifts, this may be due to the specific manner in which black hole seeding, growth, and feedback are implemented in these models.  AGN feedback is implemented in our models via two different modes: ``jet mode'' (also called ``radio mode'') and ``radiative mode'' (also called ``quasar mode'' or ``bright mode''). ``Radio mode''  feedback (\citealt{Croton2006}, \citetalias{Somerville2008}) is implemented as a heating term which (in the Santa Cruz SAMs) scales as a power-law function of the black hole mass, based on observations of radio jets in nearby galaxy groups and clusters (see \citetalias{Somerville2008}). The jets are assumed to be able to couple efficiently with the hot gas only when the cooling time is longer than the dynamical time, which tends to be the case in massive halos at late times. As a result, jet mode feedback becomes effective only in massive halos at redshifts below about $z\sim 1$.

The Santa Cruz SAMs also include ``radiative mode'' feedback, in which cold gas can be removed from galaxies via radiation pressure driven winds associated with radiatively efficient accretion onto a black hole. Although this mode can act at high redshift, it tends to have little effect on galaxy properties, because it is assumed that the winds can only act ``ejectively'' on the cold interstellar gas, and new gas tends to cool rapidly and replenish the cold gas reservoir, especially at high redshift. However, the treatment of radiative mode feedback in the existing models is based on an older suite of hydrodynamic simulations of binary mergers between idealized galaxies with no initial hot gas halo. More recent cosmological zoom-in simulations including thermal and kinetic feedback from radiatively efficient black hole accretion find that there is also a strong \emph{preventative} feedback effect, as these winds can significantly reduce the density of gas near the centers of halos and thereby suppress cooling for much longer timescales \citep{Choi2015, Choi2017, Brennan2018}. In this picture, we might expect quenching via radiative mode feedback to be more effective at higher redshifts.

\subsection{Physical processes that shape low-luminosity galaxies}

In this work, we updated the Santa Cruz SAM treatment of photoionization squelching by implementing the fitting function presented in \citet{Okamoto2008} to model heating from an photoionizing background. Our results in fig. \ref{fig:UVLF} show that the effect is negligible at all redshifts. Some studies suggest that photoionization squelching plays a  significant role in shaping the faint end of the galaxy LF, suppressing the faint end of the UV LFs in the range $M_\text{UV} \sim -12$ -- $-10$ \citep{Shapiro2004, Iliev2005, Ocvirk2016}. Some others have shown that dense clumps are extremely hard to penetrate and become ionized due to a high recombination rate. Simulations have shown that cooling in gas clumps should be fairly efficient when densities are comparable to the virial density of a halo \citep{Noh2014}. \citet{Susa2004, Susa2004a} used radiative hydrodynamic simulations with radiative transfer to show that a photoionizing background is devastating only for low-mass systems of $V_c \lesssim 20$ \kms, which are halos close to the atomic cooling limit. Similar results are also presented by \citet{Okamoto2008}, where only extremely low density gas clumps are completely evaporated and the overall effect of a photoionizing background on galaxy formation is much weaker than previously thought.

In our models, stellar-driven winds play a dominant role in shaping the slope of the faint-end of the UVLF as well as the location of the turnover. This type of simple empirical model for stellar-wind feedback has been widely adopted in both SAMs and some cosmological numerical hydrodynamic simulations. The recipe has been shown to be quite successful in reproducing observations at lower redshifts for galaxies across a wide range of masses. As discussed in \citetalias{Somerville2015}, SF in low-mass halos is self-regulated by stellar feedback. For instance, when SF becomes more efficient, more gas is ejected by energetic stellar winds, and hence reducing the supply of cold gas and yielding less efficient SF, and vice versa. This effect has also been examined and demonstrated in a number of studies \citep{Schaye2010, Haas2013, White2015a}. Here we show that this remains the case in low-mass halos up to very high redshift, even in models with metallicity dependent, \molh-based star formation. 

The free parameters in our physically motivated stellar feedback recipe have only been tuned to match observed SMF and stellar-to-halo mass ratio at $z \sim 0$. In addition to the calibrated fiducial values, we also show example cases where we systematically vary the SN feedback slope $\alpha_\text{rh}$ slightly by $\pm 0.4$ and $\pm 0.8$ while keeping the rest of the model unchanged. Our results clearly demonstrate that this parameter indeed plays a dominant role even at extreme redshifts. However, we also show that the more massive galaxies are insensitive to variations in this parameter, because stellar-driven winds cannot efficiently escape the deep potential wells of the halos that host these objects.

Our model is one of the very few simulations that is capable of resolving objects as tiny as $V_c \sim 20$ \kms in a cosmological context, close to the atomic cooling limit. We find that the turnover in the UVLF due to the atomic cooling limit occurs at around $M_\text{UV} \sim -8$ at $z\sim 4$, moving slightly brighter to about $M_\text{UV} \sim -9$ at $z\sim 10$. We find that the magnitude where the turnover occurs shifts brighter or fainter by up to $\sim 1$ magnitude under variations in the stellar feedback parameter $\alpha_\text{rh}$. This is because $\alpha_\text{rh}$ changes the slope of the median relationship between halo mass and stellar mass or SFR, shifting the rest-UV magnitude corresponding to a halo circular velocity of $V_c \sim 20$ \kms.

\subsection{Probing ultra high-redshift galaxies beyond \textit{JWST}} 

The galaxy populations predicted by our model span a wide range of luminosities and redshifts. Even though JWST will be able to detect many objects during the epoch of reionization, our models predict that there will still be a significant population of objects too faint to be detected even by \textit{JWST}.  Although these galaxies are unlikely to play a significant role in cosmic reionization, they are thought to have hosted Pop III stars and polluted the ISM with the first heavy elements.  Therefore, constraints on these objects are a very important missing piece in the current formation theory for stars and galaxies. Probing these objects directly requires instruments with sensitivity many times higher than what \textit{JWST} has achieved. However, the bulk effects of these objects can be studied via metal absorption (e.g. \citealt{Finlator2013}) or intensity mapping \citep{Visbal2010, Visbal2011}. Many ongoing and planned intensity mapping pathfinders (e.g. CHIME, HERA, HIRAX, Tianlai, BINGO, LOFAR, MeerKat, CONCERTO, STARFIRE) are paving the way to future high-$z$ large-scale multiline intensity mapping surveys. 

On the other hand, planned wide-field surveys, such as those that will be carried out with Euclid and the Wide-Field Infrared Survey Telescope (WFIRST), will probe unprecedented areas, providing  constraints on the massive, bright galaxy populations \citep{Racca2016, Spergel2015}.

We plan to exploit our model framework to forecast and provide an interpretive framework for these and other observations in future projects. We plan to further explore the progression of cosmic reionization arising from our predicted galaxy populations, and compare that to the latest observational constraints from the Ly$\alpha$ forest and the Thomson scattering optical depth for the cosmic microwave background (Yung et al., in prep).

\section{Summary and Conclusions}
\label{section:summary}

In this work, we presented predictions for galaxy populations that are expected to be detected in upcoming \textit{JWST} observations, and showed how they can  potentially constrain the physical processes that govern the formation and evolution of these objects. Our galaxies are modeled using the well-established Santa Cruz semi-analytic model with the recently updated multiphase gas partitioning and \molh-based SF recipes. We also used semi-analytic dark matter halo merger trees that are constructed based on the EPS formalism to achieve the very wide dynamic range and computational efficiency required for forecasting observable properties of galaxy populations over a wide range of masses and redshifts. Moreover, we adopted the updated Planck cosmology and the model parameters were recalibrated to match the latest observational constraints near $z\sim 0$.

By exploiting the high efficiency of our model, we were also able to systematically vary the SF recipes, as well as the sub-grid physical parameters for gas depletion timescale and stellar feedback relation slope, in a controlled manner. We use these results to explore the physical processes that have degenerate effects on galaxies, which create tension in matching galaxy properties and other cosmological observables. We also  discuss whether these processes can be disentangled and what we expect to learn from the upcoming deep-field observations. 

Predictions for rest-frame UV luminosity functions at $z = 4$ -- 10 are presented and are compared to existing observations and other models. We include the effects of dust attenuation using an empirical dust recipe with a redshift dependent dust-to-metal ratio. Although the free parameters in our model are only calibrated to match observations at $z \sim 0$, our results matches surprisingly well with UV LF constraints at $z > 4$. In addition, our results agree extremely well with previous theoretical studies, particularly for more luminous galaxies. We predicted that the faint end of the UV luminosity functions will remain steep below the current detection limit until $M_\text{UV} \sim -9$. We showed that the gas depletion time and the choice of star formation recipe have strong influences on star formation in luminous galaxies. Conversely, the effect of feedback from AGN is found to be negligible, although this may be due to shortcomings in our modeling of black hole seeding, accretion, and/or feedback.  However, starlight from the most intrinsically UV-luminous galaxies is also heavily obscured by dust, making it extremely difficult to disentangle these degenerate effects from multiple physical processes. On the other hand, star formation in low-mass halos seems to be most strongly affected by stellar feedback, with photoionization feedback having a negligible effect on the populations that we studied. 

We estimated the apparent magnitudes of our predicted galaxy population utilizing the published \textit{JWST} NIRCam broadband filters and presented them in the form of one-point distribution functions, which may be compared directly with observations. We also estimated the effects of dust attenuation and illustrated the sensitivity to stellar feedback efficiency. We show that, with a simple dust model, the effect of dust is only significant for rapidly star forming, metal rich galaxies and is more important at lower redshifts. \newline

We summarize our main conclusions below. \newline

(i) A relatively simple and computationally efficient semi-analytic model, which has been calibrated only to $z \sim 0$ observations, produces predictions that agree remarkably well with observed UV luminosity functions from $z\sim 4$--10.

(ii) Star formation physics and gas depletion time are dominant in determining the abundance of bright, massive galaxy populations, and these physical processes are degenerate with the attenuation due to dust. Therefore, it is critical to obtain independent probes on the dust content of high redshift galaxies in order to be able to disentangle the underlying physics.

(iii) The faint-end slope of UV LFs is mainly sensitive to the scaling of the mass-loading factor for stellar driven winds with halo or galaxy properties. We find that the effect of photoionization squelching on galaxies that will be detectable with \textit{JWST} is negligible. \textit{JWST} observations will be able to place important constraints on stellar feedback. However, \textit{JWST} will not be able to probe down to the atomic cooling limit.

(iv) In our models, the absolute magnitude at which the ``turnover'' in the UV LF occurs due to the atomic cooling limit is also sensitive to the adopted recipe for stellar feedback. This is because stellar feedback changes the halo mass that hosts galaxies of a given UV luminosity.

\section*{Acknowledgements}
The authors of this paper would like to thank Aldo Rodr\'iguez-Puebla, Viraj Pandya, Peter Behroozi, Takashi Okamoto, Daniel Weisz, Eli Visbal, and David Spergel for useful discussions. We also thank Kristian Finlator, Rachael Livermore, and Guochao Sun for providing unpublished results for comparison. We also thank Rebecca Sesny and Dylan Simon for the creation of the project website for online data release. AY and RSS  thank the Downsbrough family for their generous support, and gratefully acknowledge funding from the Simons Foundation.

\bibliographystyle{mnras}
\bibliography{library.bib}

\appendix

\section{Tabulated values for selected UV LFs and AMFs}
\label{section:tabulated}
\setcounter{table}{0} \renewcommand{\thetable}{A\arabic{table}}

Tabulated UV LFs from our fiducial model, with and without dust attenuation, and AMFs calculated with our nominal NIRCam F200W filter are provided in the tables \ref{table:tab_LF_dust} -- \ref{table:tab_AMF}. Other UV LFs shown in this work are available online.

\begin{table}
	\centering
	\caption{Tabulated UV LFs at $z = 4$ -- 10 from our fiducial model \emph{with} dust attenuation.}
	\label{table:tab_LF_dust}
	\begin{tabular}{cccccccccc}
		\hline
		 & \multicolumn{7}{c}{$\log_{10}(\phi\,[\text{mag}^{-1}\,\text{Mpc}^{-3}])$} \\
		$M_\text{UV}$ & $z = 4$ & $z = 5$  & $z = 6$  & $z = 7$  & $z = 8$  & $z = 9$  & $z = 10$   \\
		\hline
-23.0 & -6.40 & -6.51 & -7.04 & -7.22 & -7.46 & -8.03 & -9.00 \\
-22.5 & -5.18 & -5.24 & -5.52 & -5.94 & -6.29 & -6.87 & -7.67 \\
-22.0 & -4.32 & -4.45 & -4.71 & -5.16 & -5.65 & -6.10 & -6.68 \\
-21.5 & -3.79 & -3.93 & -4.20 & -4.51 & -4.94 & -5.46 & -6.06 \\
-21.0 & -3.44 & -3.50 & -3.70 & -4.09 & -4.44 & -4.89 & -5.49 \\
-20.5 & -2.99 & -3.16 & -3.38 & -3.63 & -3.97 & -4.53 & -5.02 \\
-20.0 & -2.72 & -2.84 & -2.99 & -3.30 & -3.67 & -4.09 & -4.63 \\
-19.5 & -2.47 & -2.57 & -2.73 & -3.03 & -3.33 & -3.75 & -4.30 \\
-19.0 & -2.24 & -2.37 & -2.53 & -2.76 & -3.05 & -3.51 & -3.97 \\
-18.5 & -2.08 & -2.16 & -2.37 & -2.62 & -2.92 & -3.24 & -3.67 \\
-18.0 & -1.95 & -2.00 & -2.19 & -2.36 & -2.70 & -2.95 & -3.44 \\
-17.5 & -1.83 & -1.91 & -2.05 & -2.21 & -2.42 & -2.76 & -3.14 \\
-17.0 & -1.69 & -1.75 & -1.88 & -2.07 & -2.30 & -2.56 & -2.93 \\
-16.5 & -1.60 & -1.65 & -1.70 & -1.88 & -2.04 & -2.32 & -2.67 \\
-16.0 & -1.49 & -1.49 & -1.57 & -1.73 & -1.92 & -2.20 & -2.49 \\
-15.5 & -1.34 & -1.34 & -1.43 & -1.58 & -1.74 & -1.96 & -2.32 \\
-15.0 & -1.22 & -1.25 & -1.28 & -1.37 & -1.61 & -1.83 & -2.09 \\
-14.5 & -1.08 & -1.11 & -1.16 & -1.30 & -1.46 & -1.64 & -1.98 \\
-14.0 & -1.04 & -1.02 & -0.98 & -1.12 & -1.24 & -1.49 & -1.70 \\
-13.5 & -0.92 & -0.86 & -0.91 & -0.99 & -1.15 & -1.29 & -1.56 \\
-13.0 & -0.77 & -0.77 & -0.75 & -0.84 & -0.99 & -1.16 & -1.42 \\
-12.5 & -0.69 & -0.68 & -0.67 & -0.71 & -0.84 & -1.05 & -1.25 \\
-12.0 & -0.60 & -0.53 & -0.56 & -0.58 & -0.65 & -0.85 & -1.08 \\
-11.5 & -0.49 & -0.41 & -0.41 & -0.43 & -0.54 & -0.72 & -0.90 \\
-11.0 & -0.46 & -0.37 & -0.28 & -0.36 & -0.38 & -0.52 & -0.73 \\
-10.5 & -0.34 & -0.22 & -0.18 & -0.20 & -0.29 & -0.43 & -0.58 \\
-10.0 & -0.27 & -0.15 & -0.09 & -0.12 & -0.19 & -0.28 & -0.45 \\
-9.5 & -0.22 & -0.07 & -0.00 & 0.07 & -0.03 & -0.13 & -0.27 \\
-9.0 & -0.20 & -0.01 & 0.02 & 0.01 & -0.04 & -0.19 & -0.37 \\
-8.5 & -0.17 & -0.13 & -0.13 & -0.10 & -0.21 & -0.32 & -0.45 \\
-8.0 & -0.28 & -0.20 & -0.18 & -0.20 & -0.28 & -0.40 & -0.59 \\
		\hline
	\end{tabular}
\end{table}

\begin{table}
	\centering
	\caption{Tabulated UV LFs at $z = 4$ -- 10 from our fiducial model \emph{without} dust attenuation.}
	\label{table:tab_LF_nodust}
	\begin{tabular}{cccccccccc}
		\hline
		& \multicolumn{7}{c}{$\log_{10}(\phi\,[\text{mag}^{-1}\,\text{Mpc}^{-3}])$} \\
		$M_\text{UV}$ & $z = 4$ & $z = 5$  & $z = 6$  & $z = 7$  & $z = 8$  & $z = 9$  & $z = 10$   \\
		\hline
-23.0 & -3.39 & -3.68 & -4.19 & -4.61 & -5.25 & -5.92 & -6.81 \\
-22.5 & -3.19 & -3.48 & -3.86 & -4.31 & -4.87 & -5.50 & -6.26 \\
-22.0 & -3.03 & -3.27 & -3.63 & -4.01 & -4.58 & -5.16 & -5.86 \\
-21.5 & -2.93 & -3.10 & -3.39 & -3.79 & -4.25 & -4.81 & -5.47 \\
-21.0 & -2.74 & -2.94 & -3.22 & -3.59 & -4.00 & -4.56 & -5.14 \\
-20.5 & -2.67 & -2.78 & -2.98 & -3.35 & -3.75 & -4.18 & -4.78 \\
-20.0 & -2.53 & -2.65 & -2.83 & -3.15 & -3.55 & -3.97 & -4.47 \\
-19.5 & -2.38 & -2.54 & -2.72 & -2.98 & -3.29 & -3.70 & -4.24 \\
-19.0 & -2.30 & -2.35 & -2.56 & -2.78 & -3.04 & -3.48 & -3.92 \\
-18.5 & -2.08 & -2.18 & -2.32 & -2.57 & -2.92 & -3.18 & -3.65 \\
-18.0 & -2.02 & -2.04 & -2.22 & -2.36 & -2.64 & -2.93 & -3.40 \\
-17.5 & -1.84 & -1.93 & -2.03 & -2.23 & -2.44 & -2.77 & -3.12 \\
-17.0 & -1.72 & -1.74 & -1.89 & -2.04 & -2.28 & -2.56 & -2.92 \\
-16.5 & -1.62 & -1.64 & -1.71 & -1.90 & -2.06 & -2.31 & -2.65 \\
-16.0 & -1.48 & -1.50 & -1.56 & -1.73 & -1.92 & -2.20 & -2.51 \\
-15.5 & -1.33 & -1.34 & -1.44 & -1.58 & -1.75 & -1.97 & -2.31 \\
-15.0 & -1.21 & -1.26 & -1.29 & -1.37 & -1.61 & -1.83 & -2.09 \\
-14.5 & -1.09 & -1.12 & -1.16 & -1.31 & -1.46 & -1.64 & -1.98 \\
-14.0 & -1.05 & -1.01 & -0.99 & -1.13 & -1.24 & -1.49 & -1.70 \\
-13.5 & -0.92 & -0.86 & -0.91 & -0.99 & -1.15 & -1.29 & -1.56 \\
-13.0 & -0.77 & -0.77 & -0.75 & -0.84 & -0.99 & -1.16 & -1.42 \\
-12.5 & -0.69 & -0.68 & -0.67 & -0.71 & -0.84 & -1.05 & -1.25 \\
-12.0 & -0.60 & -0.53 & -0.56 & -0.58 & -0.65 & -0.85 & -1.08 \\
-11.5 & -0.49 & -0.41 & -0.41 & -0.43 & -0.54 & -0.72 & -0.90 \\
-11.0 & -0.46 & -0.37 & -0.28 & -0.36 & -0.38 & -0.52 & -0.73 \\
-10.5 & -0.34 & -0.22 & -0.18 & -0.20 & -0.29 & -0.43 & -0.58 \\
-10.0 & -0.27 & -0.15 & -0.09 & -0.12 & -0.19 & -0.28 & -0.45 \\
-9.5 & -0.22 & -0.07 & -0.00 & 0.07 & -0.03 & -0.13 & -0.27 \\
-9.0 & -0.20 & -0.01 & 0.02 & 0.01 & -0.04 & -0.19 & -0.37 \\
-8.5 & -0.17 & -0.13 & -0.13 & -0.10 & -0.21 & -0.32 & -0.45 \\
-8.0 & -0.28 & -0.20 & -0.18 & -0.20 & -0.28 & -0.40 & -0.59 \\
		\hline
	\end{tabular}
\end{table}

\begin{table}
	\centering
	\caption{Tabulated AMFs at $z = 4$ -- 10 from our fiducial model \emph{with} dust attenuation calculated using our nominal NIRCam F200W filter.}
	\label{table:tab_AMF}
	\begin{tabular}{cccccccccc}
		\hline
		& \multicolumn{7}{c}{$\log_{10}(\phi\,[\text{mag}^{-1}\,\text{Mpc}^{-3}])$} \\
		$m$ & $z = 4$ & $z = 5$  & $z = 6$  & $z = 7$  & $z = 8$  & $z = 9$  & $z = 10$   \\
		\hline
28.0 & -1.99 & -2.26 & -2.58 & -2.89 & -3.29 & -3.76 & -4.40 \\
28.5 & -1.87 & -2.13 & -2.40 & -2.70 & -3.02 & -3.52 & -4.04 \\
29.0 & -1.71 & -2.01 & -2.25 & -2.50 & -2.90 & -3.24 & -3.72 \\
29.5 & -1.63 & -1.85 & -2.07 & -2.32 & -2.62 & -2.96 & -3.50 \\
30.0 & -1.49 & -1.73 & -1.92 & -2.13 & -2.41 & -2.78 & -3.21 \\
30.5 & -1.35 & -1.59 & -1.76 & -1.99 & -2.25 & -2.56 & -2.96 \\
31.0 & -1.25 & -1.44 & -1.61 & -1.81 & -2.04 & -2.34 & -2.74 \\
31.5 & -1.12 & -1.34 & -1.46 & -1.66 & -1.90 & -2.20 & -2.54 \\
32.0 & -1.02 & -1.19 & -1.32 & -1.53 & -1.72 & -1.98 & -2.38 \\
32.5 & -0.94 & -1.10 & -1.22 & -1.32 & -1.60 & -1.86 & -2.13 \\
33.0 & -0.81 & -0.97 & -1.04 & -1.27 & -1.43 & -1.65 & -2.00 \\
33.5 & -0.70 & -0.84 & -0.94 & -1.09 & -1.24 & -1.50 & -1.78 \\
34.0 & -0.60 & -0.76 & -0.80 & -0.93 & -1.12 & -1.31 & -1.57 \\
34.5 & -0.50 & -0.62 & -0.68 & -0.78 & -1.00 & -1.18 & -1.50 \\
35.0 & -0.45 & -0.51 & -0.62 & -0.68 & -0.80 & -1.05 & -1.29 \\
35.5 & -0.36 & -0.40 & -0.44 & -0.51 & -0.67 & -0.87 & -1.11 \\
36.0 & -0.27 & -0.32 & -0.32 & -0.43 & -0.52 & -0.77 & -0.95 \\
36.5 & -0.19 & -0.22 & -0.23 & -0.29 & -0.39 & -0.57 & -0.78 \\
37.0 & -0.16 & -0.09 & -0.10 & -0.16 & -0.28 & -0.43 & -0.65 \\
37.5 & -0.15 & -0.06 & -0.03 & -0.07 & -0.19 & -0.30 & -0.47 \\
38.0 & -0.23 & 0.02 & 0.07 & 0.10 & -0.01 & -0.13 & -0.32 \\
		\hline
	\end{tabular}
\end{table}

\section{Re-calibration for Planck Cosmology}
\label{section:recalibration}
\setcounter{figure}{0} \renewcommand{\thefigure}{B\arabic{figure}}

\begin{figure*}
	\includegraphics[width=2.0\columnwidth]{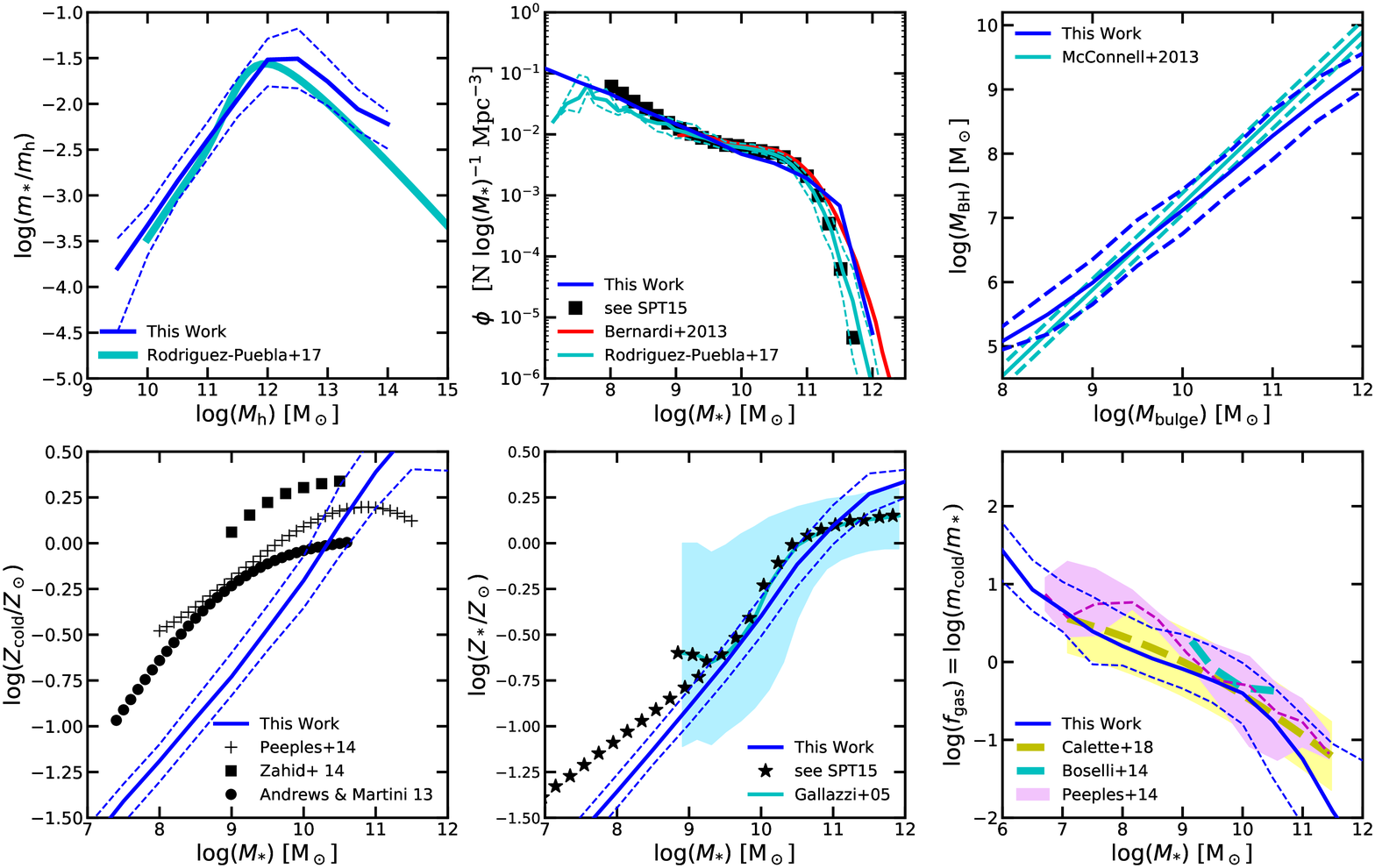}
	\caption{Outputs at $z = 0$ from our fiducial model compared to the observational constraints on \textit{top row from left to right}: stellar-to-halo mass ratio, stellar mass function, and $M_\text{BH}$-$M_\text{bulge}$ relation; \textit{bottom row from left to right}: cold gas metallicity, stellar metallicity, and gas fraction reported by various studies. See text for full details.}
	\label{fig:calibration}
\end{figure*}

\begin{figure}
	\includegraphics[width=\columnwidth]{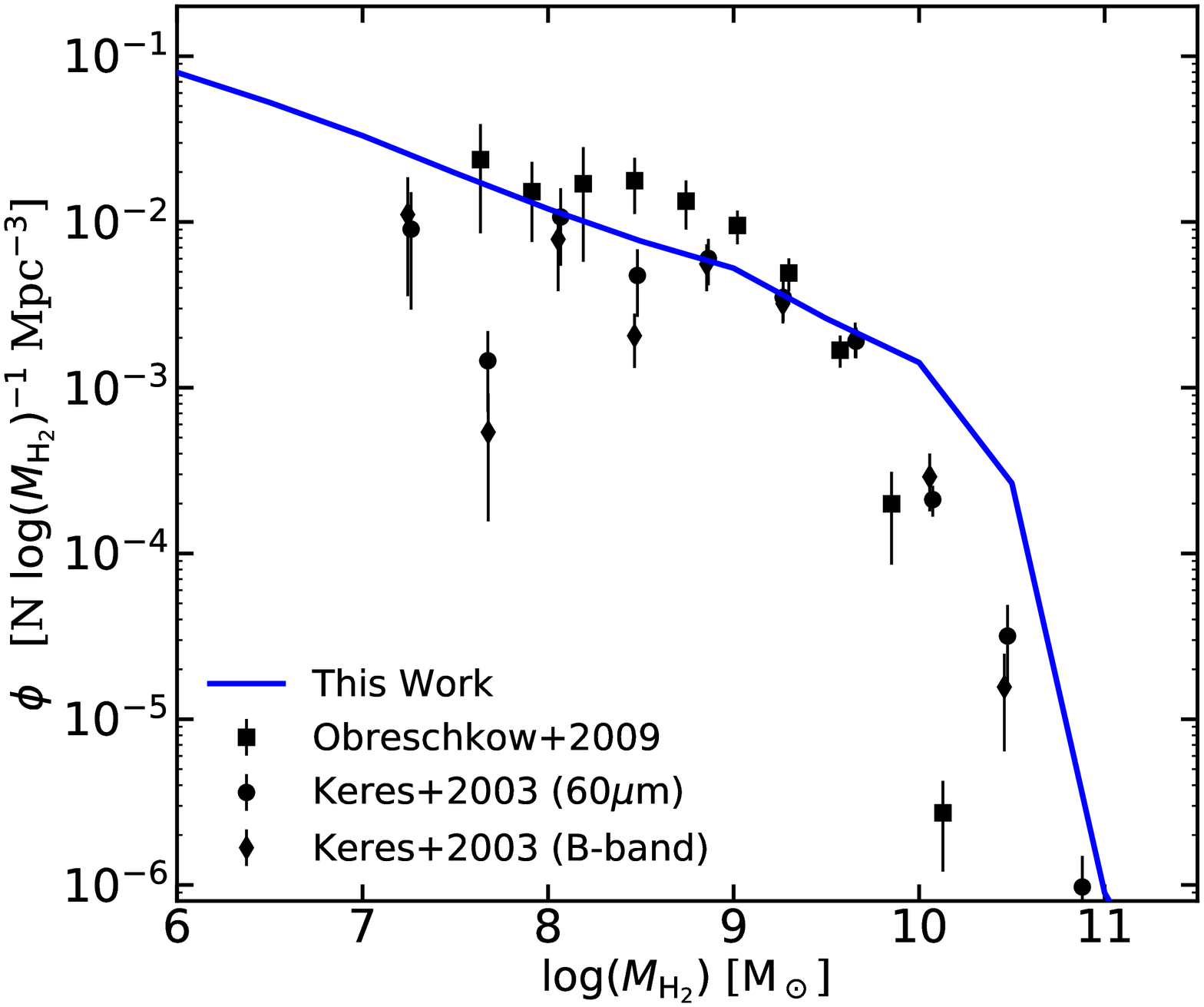}
	\caption{\molh\ mass function at $z = 0$ from our fiducial model compared to the observational constraints presented in \citet{Keres2003} and \citet{Obreschkow2009}.}
	\label{fig:H2_crosscheck}
\end{figure}

In this appendix, we briefly summarize the calibration of our model after updating the cosmological parameters to values consistent with recent constraints from Planck. The free parameters in our model are calibrated by hand such that the outputs from our fiducial model match observations at $z \sim 0$. The calibration quantities are the stellar-to-halo mass ratio and stellar mass function, stellar mass-metallicity relation, cold gas fraction versus stellar mass relation for disk-dominated galaxies, and the black hole mass vs. bulge mass relation. We adopt observational constraints for these quantities from \citet[][and references therein]{Rodriguez-Puebla2017}, \citet{Bernardi2013}, \citet{Gallazzi2005}, \citet{Peeples2014}, \citet{Calette2018}, and \citet{McConnell2013}. We show our model outputs compared with these observational constraints in fig. \ref{fig:calibration}. In addition, we show two quantities used as a cross-check but not used directly in the calibration: cold gas phase mass-metallicity relation (lower-left panel in fig. \ref{fig:calibration}) and the \molh\ mass function (fig. \ref{fig:H2_crosscheck}), along with observational constraints from \citet{Obreschkow2009}, \citet{Keres2003}, \citet{Andrews2013}, \citet{Zahid2013}, and \citet{Boselli2014}. For scaling relations, the blue solid line marks the median and the dashed lines mark the 16\textsuperscript{th} and 84\textsuperscript{th} percentile for central galaxies. Both central and satellite galaxies are included in the distribution functions (such as the SMF and \molh\ mass function). \citetalias{Somerville2015} has shown that the different SF recipes produce results that converge at $z \sim 0$. The re-calibration for updated cosmology have uniform effects across the models and therefore should not qualitatively change the results among models. Hence, we refer reader to such work for the relative differences among SF models at low redshifts.  

Traditionally, there has been tension in the predicted physical properties in galaxy formation models. In one case, matching the stellar fraction closely tends to lead to an excess in gas fraction. Since the underlying star formation efficiency that has immediate effects on stellar fraction is directly related to the production of metals, which our multiphase-gas partitioning, star formation, and dust attenuation recipes are extremely sensitive to, calibrating our model to simultaneously match the observed gas fraction, stellar metallicity, and stellar fraction is very challenging. The uncertainty of about a factor of two in the observed normalization of the Kennicutt SF law and the observed gas fractions provides us with some leeway for calibrating our model. After carrying out multiple tests with different configurations, we find that leaving the SF timescale $\tau_{*,0}$ close to unity yields the best results, with gas fractions slightly higher than values reported by \citet{Calette2018}. If we strictly enforce the gas fraction to match observations by decreasing $\tau_{*,0}$, we will also need to compensate for that by increasing AGN feedback to keep the model from overproducing massive galaxies. 

While we configure the SF timescale and AGN feedback to fit the massive end of the stellar-to-halo mass ratio and stellar mass function, the faint populations seem to be more sensitive to the SN feedback slope $\alpha_\text{rh}$. Structure forms earlier in the Planck cosmology relative to the WMAP-5 cosmology used in our previous work, leading to a higher number density of low-mass dark matter halos at early times. This requires us to  increase $\alpha_\text{rh}$ to suppress the formation of low-mass galaxies in order to match both the stellar-to-halo mass ratio and stellar mass function, which comes with a side effect of further steepening the mass-metallicity relation. As discussed in previous works \citetalias[e.g.][]{Somerville2015}, this tension, along with other discrepancies at intermediate redshifts \citep{White2015a} hint that our rather simple recipe for stellar feedback needs to be revised. Moreover, since the abundance of low-mass galaxies is very sensitive to the choice of $\alpha_\text{rh}$, deviating from the calibrated value would certainly would certainly raise tension with observational constraints at $z \sim 0$. The impact of alternative $\alpha_\text{rh}$ values at $z \sim 0$ has been examined in \citep{White2015a}, which they found that making alternate assumptions for ejected mass would lead to a $\sim 0.25$ dex changes relative to the fiducial model. Some of the values, including the ones disfavored by local observations are explored in this work in effort to quantify its effect on the low-mass galaxy populations. The differences among the three star formation recipes presented in this work have been explored in \citetalias{Somerville2015}. 

\section{Testing the halo mass function}
\setcounter{figure}{0} \renewcommand{\thefigure}{C\arabic{figure}}
In this work, we use the fitting functions for halo mass functions provided by \citet{Rodriguez-Puebla2016} that is fitted to the Bolshoi-Planck simulations \citep{Klypin2016}. However, the mass resolution from Bolshoi-Planck simulation is well above $V_\text{vir} \sim 20$ \kms which our grid of root halos reaches. To check the validity of these fitting functions for HMF, we used are compared to very high resolution, small box simulations that are similar to the ones presented in \citet{Visbal2018}. A comparison between the halo mass functions and the numerical simulations is shown in fig. \ref{fig:hmf_app}.

\begin{figure}
	\includegraphics[width=\columnwidth]{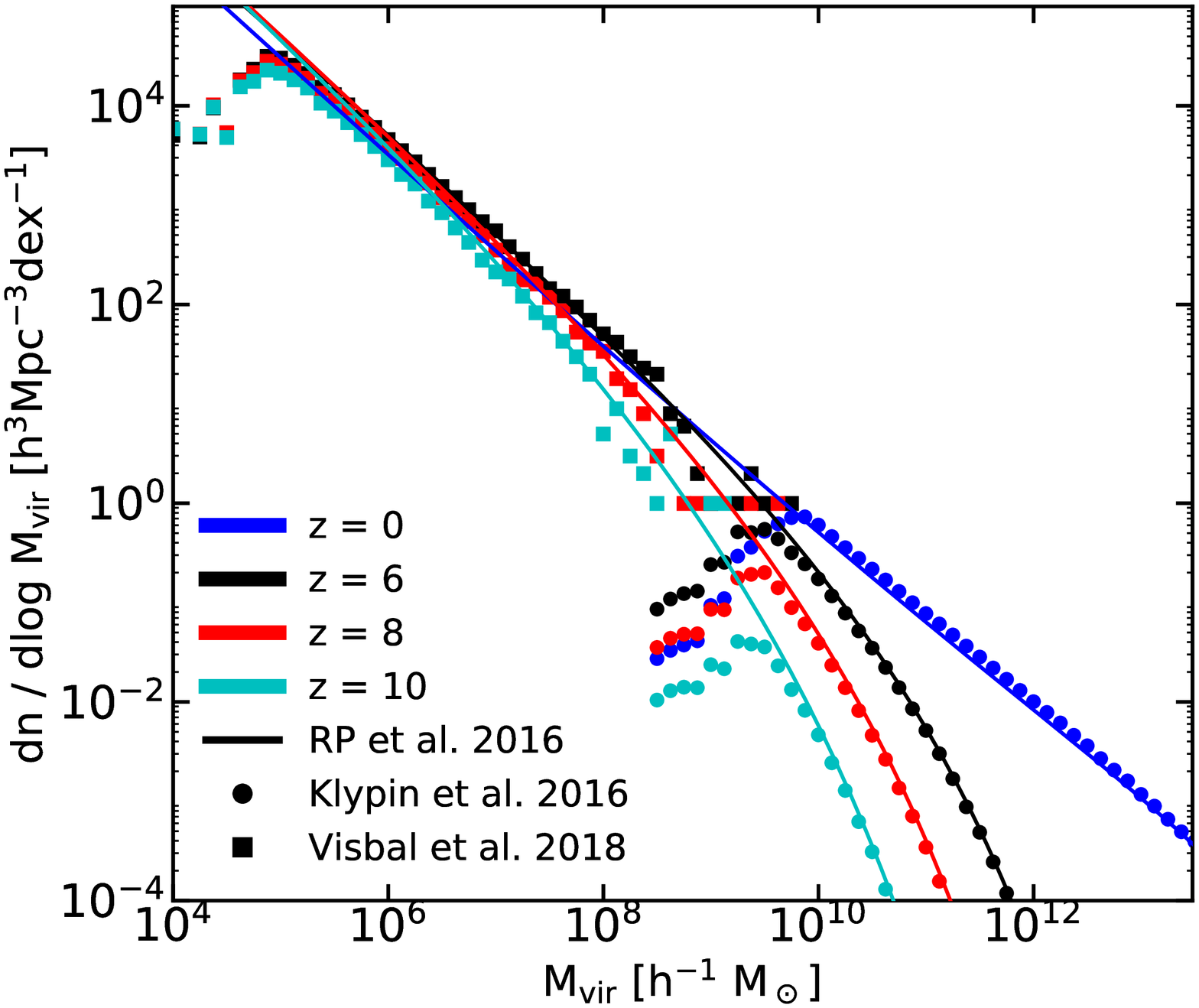}
	\caption{A comparison of the HMF fitting functions from \citet[solid line]{Rodriguez-Puebla2016} to numerical simulations from \citet[circle marker]{Klypin2016} and \citet[square marker]{Visbal2018}.Plot elements are color-matched by redshift. }
	\label{fig:hmf_app}
\end{figure}

\bsp
\label{lastpage}
\end{document}